\documentclass[11pt]{article}
\pdfoutput=1
\usepackage{mathtools}

\usepackage{setspace}
\usepackage{lipsum}
 \usepackage{multirow}
\usepackage{fullpage}
\usepackage{amsmath}
\usepackage{amssymb}
\usepackage{setspace}
\usepackage{bbm}
\usepackage{dsfont}
\usepackage{graphics}
\usepackage{longtable}
\usepackage[font=footnotesize,labelfont=bf,justification=centerlast,width=.9\textwidth]{caption}
\usepackage{color}
\usepackage{etoolbox}
 \usepackage{multirow}
\usepackage{booktabs}
\usepackage{array}
\usepackage{hyperref}
\usepackage{cite}
\usepackage[normalem]{ulem}
\hypersetup{
    bookmarks=true,%
    colorlinks,%
    citecolor=blue,%
    filecolor=black,%
    linkcolor=black,%
    urlcolor=blue
}

\newcommand{\twobytwo}[4]{\left(\begin{array}{cc}#1&#2\\#3&#4\end{array}\right)}
\newcommand\rref[1]{(\ref{#1})}
\newcommand{\be}{\begin{equation}}
\newcommand{\ee}{\end{equation}}
\newcommand{\bes}{\begin{equation*}}
\newcommand{\ees}{\end{equation*}}
\newcommand{\bea}{\begin{eqnarray}}
\newcommand{\eea}{\end{eqnarray}}
\newcommand{\beas}{\begin{eqnarray*}}
\newcommand{\eeas}{\end{eqnarray*}}

\newcommand{\lns}{l_{\rm NS}}

\newcommand{\bmat}{\begin{bmatrix}}
\newcommand{\emat}{\end{bmatrix}}

\def\le{\left}
\def\ri{\right}

\def\z{\mathfrak{z}}
\newcommand{\CC}{\mathbb{C}} 

\newcommand{\QQ}{\mathbb{Q}} 
 
\newcommand{\ZZ}{\mathbb{Z}}

\newcommand{\Tr}{{\rm {Tr}}}
\newcommand{\ima}{\text{Im}}

\renewcommand{\H}{\mathbb{H}}
\newcommand{\Z}{\mathbb{Z}}
\newcommand{\tr}{ {\rm Tr}}

\def\eg{{\it e.g.~}}

\begin{document}
\numberwithin{equation}{section}
{
\begin{titlepage}
\begin{center}

\hfill \\
\hfill \\
\vskip 0.75in

{\Large \bf  Siegel Paramodular Forms and Sparseness in AdS$_3$/CFT$_2$}\\

\vskip 0.4in

{\large Alexandre Belin${}^a$, Alejandra Castro${}^a$, Jo\~{a}o Gomes${}^{a}$, and Christoph A.~Keller${}^{b,c}$}\\
\vskip 0.3in

${}^{a}${\it Institute for Theoretical Physics, University of Amsterdam,
Science Park 904, Postbus 94485, 1090 GL Amsterdam, The Netherlands} \vskip .5mm
${}^{b}${\it Department of Mathematics, ETH Zurich, Raemistrasse 101, \\ CH-8092 Zurich, Switzerland}
\vskip .5mm
${}^{c}${\it Department of Mathematics, University of Arizona, 617 N. Santa Rita Ave., \\
Tucson, AZ 85721-0089, U.S.A.} \vskip .5mm
%\textcolor{red}{Check info above}\\
\texttt{a.m.f.belin@uva.nl, a.castro@uva.nl, j.m.vieiragomes@uva.nl, christoph.keller@math.ethz.ch}

\end{center}

\vskip 0.35in

\begin{center} {\bf ABSTRACT } \end{center}
We discuss the application of Siegel paramodular forms to the counting of polar states in symmetric product orbifold CFTs. We present five special examples and provide exact analytic counting formulas for their polar states. The first example reproduces the known result for  type IIB supergravity on  AdS$_3\times S^3\times K3$, whereas  the other {four examples} give new counting formulas. Their crucial feature is that the low energy spectrum is very sparse, which suggests the existence of a suitable dual supergravity theory.  These examples open a path to novel realizations of AdS$_3$/CFT$_2$. \vfill

\noindent \today

\end{titlepage}
}
%%%%%%%%%%%%%%%%%%%%%%%%%%%%%%%%%%%%%%%%%%%%%%%%%%%%%%%%%%%%%%%%%%

%%%%%%%%%%%%%%%%%%%%%
\newpage

\tableofcontents

\section{Introduction}

The AdS/CFT correspondence imposes strong requirements on holographic CFTs.
One central requirement is that they capture the entropy of black holes through microstate counting. Modular invariance has from the very beginning been a crucial ingredient for this. It provided the first pivotal insight into black hole microstates \cite{Strominger:1996sh}, which lead to an universal holographic explanation of the Bekenstein-Hawking entropy \cite{Strominger:1997eq}. More generally, in AdS$_3$ quantum gravity it restricts rather dramatically a Euclidean gravitational path integral \cite{Dijkgraaf:2000fq,Witten:2007kt,Maloney:2007ud,CastroGaberdielHartmanEtAl2012,KellerMaloney2015}. In conformal field theory, the exploitation of modular properties, among other features, gives stringent conditions on the spectrum of the theory via the modular bootstrap \cite{Hellerman2011,FriedanKeller2013,QuallsShapere2014,Anous:2018hjh}, and general constraints on their holographic fitness  has been investigated in  \cite{Keller:2011xi,BenjaminChengKachruEtAl2016,Belin:2016yll}. 

For holographic CFTs there are constraints not just on  black hole states, but also on perturbative ones, that is light states. 
The growth of the light spectrum can serve as an important diagnostic tool for finding possible holographic duals: on the one hand,
it is closely linked to how far the Cardy regime extends \cite{Hartman:2014oaa}, namely if we can expect the usual Bekenstein-Hawking entropy also for black holes at temperatures of order one. On the other hand, it allows us to identify the low energy theory. The growth of light states encodes important properties of the gravity dual, for example whether the theory is local on AdS scales or not \cite{Belin:2016dcu}.
If the growth is Hagedorn, that is the entropy is linear in the energy $h$,
\be\label{eq:cH1}
\log \rho(h) \sim h \ ,
\ee
this indicates that the theory is in a stringy-like regime. Moreover, if the coefficient multiplying $h$ in (\ref{eq:cH1}) is not bigger than $2\pi$, then the theory still has the same Hawking-Page transition, which is why \cite{Hartman:2014oaa} called this a \emph{sparse} spectrum. If the number of states grows even slower, namely if the entropy goes like a smaller power of the weight,
\be\label{eq:cS1}
\log \rho(h) \sim h^{\alpha} ~,\qquad \alpha < 1\ , 
\ee
then we will call such a spectrum \emph{very sparse}. In particular such a spectrum
indicates that the dual theory is in a supergravity-like regime: for a local field theory in $d$ dimensions we would indeed expect $\alpha = (d-1)/d$. We are interested in such very sparse spectra, and our goal is to find new examples of them.

Note that also from a modular point of view the light spectrum is more interesting than the heavy spectrum. The Cardy formula that captures the entropy of black holes is completely universal in the sense that it is fixed by modular invariance: any CFT will have Cardy growth for sufficiently high energies. The light spectrum on the other hand is not fixed by modular invariance. It is therefore not universal, and there are potentially a great many different possibilities allowed.

Our aim is to identify and quantify CFTs with slow growing perturbative spectra.
We will propose {new} examples of  counting formulas where the growth of light states is very sparse. 
To this end we consider a family of generalized partition functions $\chi_m(\tau,z)$, which by their nature have good modular transformation properties --- mathematically they are a type of Jacobi forms. Here $m$ parametrizes the central charge.
We assemble them into a generating function,
\be \label{eq:z1}
\mathcal{Z}(\rho,\tau,z)=\sum_{m\geq0}\, \chi_m(\tau,z)\,e^{2\pi i \rho m}~.
\ee
We want to consider special families of $\chi_m$ for which (\ref{eq:z1}) becomes essentially a \emph{Siegel paramodular form}: that is, after multiplying the generating function by some prefactor, it becomes symmetric under the exchange of $\rho$ and $\tau$\footnote{There are known cases where this symmetry can be physically realized in the gravitational system, for example in the context of dyonic black holes where it corresponds to electric-magnetic duality.}. Together with the modular transformation properties of $\chi_m(\tau,z)$, the resulting form is invariant under a so-called paramodular group $\Gamma_t^+$. 
This enhanced symmetry allows us to efficiently extract the degeneracies: in particular it implies that all zeros and poles are arranged in so-called Humbert surfaces, which are images under the paramodular group. This in turns allows us to use the methods of \cite{Sen:2011mh} to extract the coefficients of the form.

The generating function (\ref{eq:z1}) of course only has this enhanced symmetry for very special families of $\chi_m$. To construct such objects, we will use a so-called exponential lift \cite{Gritsenko:1996tm,Gritsenko:1996kt,Gritsenko:1999fk}, which, starting from a seed partition function $\chi(\tau,z)$, produces such a Siegel paramodular form. From the CFT point of view, this exponential lift is essentially the generating functional of  symmetric orbifold theories. Once we know the seed, we know the paramodular transformations and divisors of the Siegel modular form.

In previous work \cite{Belin:2016knb} we investigated the growth behavior of the Fourier coefficients of such Siegel paramodular forms for heavy states: that is, when writing the coefficients of the reciprocal of the Siegel paramodular form $\Phi$ as
\be\label{eq:c00}
{1\over \Phi(\rho,\tau,z)}=\sum_{m,n,l} c(m,n,l) \, e^{2\pi i \rho m+2\pi i \tau n + 2\pi i z l }~,
\ee
we define the discriminant of a state as $\Delta = 4nm-l^2$. For heavy states with  $\Delta \gg 1$, we showed that
\be\label{eq:c11}
c(m,n,l) \approx e^{\pi \sqrt{\Delta}/t}~.
\ee
This growth corresponds to Cardy growth  in CFTs\cite{cardyformula}. 
In holography, the regime of large positive discriminant is naturally associated to a black hole regime.
We found that among the paramodular forms those having poles only at a certain Humbert surface $H_1(1)$ stood out: in that case (\ref{eq:c11}) is valid in an even larger regime than \cite{Hartman:2014oaa}.

Motivated by this extended Cardy regime, we will focus on such Siegel paramodular forms here. This will give us  five examples to investigate. One example is well known in string theory: in that case $\mathcal{Z}(\rho,\tau,z)$ is the generating function for BPS states in the D1-D5 CFT with target space $K3$ \cite{Kawai1994,Eguchi1989} and the associated Siegel modular is the reciprocal of the Igusa Cusp form \cite{Dijkgraaf:1996it}. The other four examples are new. Our focus is to understand the density of states for $\Delta<0$: this is the relevant regime for light operators that can be compared with a perturbative supergravity regime.  To count these states, we apply the ideas in \cite{Sen:2011mh}, which focused on counting negative discriminant states for the Igusa cusp form (and related counting formulas for CHL compactifications), to our new examples.

Let us briefly comment on the connection of exponentially lifted Siegel paramodular forms to symmetric orbifolds. A
symmetric product orbifold is constructed by taking $N$ copies of a CFT $\mathcal{C}$ and orbifolding by the symmetric group $S_N$:
\be
\mathcal{C}_N \equiv \frac{\mathcal{C}^{\otimes N}}{S_N} \,.
\ee
By orbifolding, we are left with a finite number of low dimensions operators in the large $N$ limit. Counting all these operators, that is taking the standard partition function, 
we find Hagedorn growth \cite{Keller:2011xi,Hartman:2014oaa}
\be\label{eq:symOrb}
\rho(h) \approx e^{2\pi h}\ . 
\ee
Here $h$ stands for the conformal dimension of the operator, $N$ plays the role of $m$ in \rref{eq:z1}, and (\ref{eq:symOrb}) holds for light states, that is states with $h \ll N$. Note that this result is completely independent of the seed theory $\mathcal{C}$.

What we have said so far would seem to indicate that symmetric orbifolds are not a good place to look for very sparse spectra. Neither are more general permutation orbifolds \cite{Haehl:2014yla,BelinKellerMaloney2016}, which do not give a growth like (\ref{eq:cS1}) \cite{Belin:2014fna} either. Indeed, the correct interpretation is probably that they describe holographic duals far from the supergravity regime. In fact, their correlators also behave very differently than a theory described by supergravity \cite{Belin:2017jli}. A way around this is to not consider symmetric orbifolds as such, but rather to deform them. This corresponds to moving around on the moduli space of the theory. To actually compute the spectrum of such a deformed theory is difficult, which is why we will not pursue this route. Instead we will concentrate on counting operators that are protected under deformations. 

In a CFT with say $N=(2,2)$ supersymmetry, such operators are for instance BPS operators. The generalized partition function or index that captures them is the elliptic genus,
defined as\cite{Witten1987,Eguchi1989,Kawai1994}
\be\label{eq:ellipgenus}
\chi(\tau,z)=\tr_{RR}\le((-1)^{F} q^{L_0-\frac{c}{24}} y^{ J_0 } \bar{q}^{\bar{L}_0-\frac{\bar{c}}{24}} \ri)\,.
\ee
Here $L_0$ and $\bar L_0$ are the left and right Virasoro generators, the fermion number is defined as $F=J_0-\bar J_0$ and $J_0,\bar J_0$ are respectively a left and right moving $R$-symmetry generator. 
The trace (\ref{eq:ellipgenus}) is defined in the Ramond sector where fermions have periodic boundary conditions. To reach perturbative states, we will therefore have to perform a spectral flow transformation.
The elliptic genus is holomorphic and has nice modular properties: it is what is called a weak Jacobi form. For fixed central charge, the space of weak Jacobi forms that can serve as elliptic genera is finite dimensional, which will allow us to  organize our search.

Physically, the advantage of studying the elliptic genus is that it is protected under deformations, which implies we can count operators at the symmetric orbifold point and still obtain a result that is valid at the supergravity point, even though it is far away on the moduli space.
It turns out that the growth for the elliptic genus can be much slower than (\ref{eq:symOrb}): in some cases, the minus sign in (\ref{eq:ellipgenus}) leads to enough cancellations that we get a growth of the form (\ref{eq:cS1}) instead.
This is in fact what happens in the D1-D5 system, which is what allows the matching to the supergravity spectrum in \cite{deBoer:1998us}. 
However, for higher central charges we generically get Hagedorn growth \cite{Benjamin:2015vkc}. Only for some special choices of seed theories do we get similar cancellations.\footnote{In \cite{Benjamin:2015vkc} a class of weak Jacobi forms leading to supergravity growth were called very special. We discuss their relation to our examples in appendix~\ref{app:veryspecial}.}

Our goal is to find counting formulas that are special enough such that the BPS states are very sparse. Our main result is that we find five of these functionals that meet this criteria. One of those examples is the D1-D5 system mentioned above, whereas our other four examples are genuinely new in this context. We take them as novel candidates for  AdS$_3$/CFT$_2$.

The paper is organized as follows. In Sec.\ \ref{sec:param}, we review Siegel paramodular forms. In Sec.\ \ref{sec:five} we construct our five examples. Sec.\ \ref{sec:method} describes how to extract negative discriminant states from our examples. Sec.\ \ref{sec:sfl} discusses spectral flow and the relation to symmetric orbifolds. In Sec.\ \ref{sec:sugra} we discuss to what extent we can interpret our results from a supergravity perspective. We conclude and discuss future directions in Sec.\ \ref{sec:discussion}.

\section{Paramodular forms}\label{sec:param}

In this section we will summarise the key features of the paramodular forms that arise from an exponential lift. The presentation here is mostly based on \cite{9781468491647,Gritsenko:1999fk,MR1616929}.

\subsection{Basic definitions}

Our starting point is to consider generating functions which are of the form
\begin{align}\label{eq:s10}
{\Phi}(\rho,\tau,z)&= \sum_{m,n,l} c_m(n,l)p^m q^n y^l 
=\sum_{m} \varphi_{k,m}(\tau,z) p^m~,
\end{align}
where 
\begin{align}
p=e^{2\pi i \rho}~,\quad q=e^{2\pi i\tau}~, \quad y=e^{2\pi iz}~,
\end{align}
with $(\tau,z,\rho)$ complex variables, and 
\be\label{eq:vp}
\varphi_{k,m}(\tau,z)=\sum_{n\geq 0,l\in \ZZ} c_m(n,l) q^n y^l~. 
\ee
We leave the range of $(m,n,l)$  in \eqref{eq:s10} and \eqref{eq:vp} unspecified and it will be narrowed as needed.  

In our discussion, $\varphi_{k,m}(\tau,z)$ transforms like a \emph{Jacobi form} \cite{9781468491647}, where $k$ is the {weight} and $m$ is the {index}.   This means $\varphi_{k,m}(\tau,z)$ is a holomorphic function on $\H\times\CC\rightarrow\CC$ that has the following transformation properties: first, under modular transformations
\be\label{eq:jf1}
\varphi_{k,m}\le({a\tau+b\over c\tau +d},{z\over c\tau +d}\ri)= (c\tau +d)^k\exp\le({2\pi i m c z^2\over c\tau +d}\ri)\varphi_{k,m}(\tau,z)~,\quad  \forall \twobytwo{a}{b}{c}{d} \in SL(2,\ZZ)~,
\ee
and second, under elliptic translations
\be\label{eq:jf2}
\varphi_{k,m}\le(\tau,{z+ \lambda \tau +\mu}\ri)= \exp\le(-{2\pi i m (\lambda^2\tau+2\lambda z +\mu)}\ri)\varphi_{k,m}(\tau,z)~, \quad \lambda,\mu \in \ZZ~.
\ee
$\Phi$ then of course inherits similar transformations properties from those characteristic of $\varphi_{k,m}$.

Returning to \eqref{eq:s10}, we are interested in \emph{Siegel paramodular forms}, whose defining feature is that in addition to transformation properties \eqref{eq:jf1}-\eqref{eq:jf2} they have an exchange symmetry:
\be\label{eq:s11}
{\Phi}(\rho,\tau,z)= {\Phi}(t^{-1}\, \tau, t {\rho},z)~.
\ee
where $t\in \ZZ^+$. This symmetry, together with the transformation properties of $\varphi_{k,m}(\tau,z)$, define the paramodular group.
Let us be more precise. We define
\be\label{Omegadef}
\Omega = \left(\begin{array}{cc}\tau&z\\z&\rho\end{array}\right)\ .
\ee
The Siegel upper half plane $\H_2$ is given by
\be\label{eq:uhp}
\det(\text{Im}(\Omega)) > 0 \ , \qquad \Tr (\text{Im}(\Omega)) > 0\ ,
\ee
and $\Phi(\Omega)$ is holomorphic on this domain. 
The paramodular group $\Gamma_t$ of level $t$  is defined as \cite{MR2208781}
\be\label{eq:defg2}
\Gamma_t :=\left[\begin{array}{cccc} 
	\Z & t\Z &\Z&\Z\\
	\Z &\Z&\Z&t^{-1}\Z\\
	\Z& t\Z&\Z&\Z\\
	t\Z&t\Z&t\Z&\Z
\end{array}\right] \cap Sp(4,\QQ)~.
\ee
It has an extension
\be\label{eq:defgammat}
\Gamma_t^+ = \Gamma_t \cup \Gamma_t V_t\ , \qquad
V_t = \frac{1}{\sqrt{t}}
\left(\begin{array}{cccc} 
	0&t&0&0\\
	1&0&0&0\\
	0&0&0&1\\
	0&0&t&0
\end{array}\right)\ .
\ee
Given a matrix $\gamma \in \Gamma_t^+$,  which we decompose into $2\times 2$ matrices as
\be
\gamma=\twobytwo{{\bf A}}{{\bf B}}{{\bf C}}{{\bf D}}~,
\ee
the action of $\gamma$ on $\Omega$ is given by
\be
\gamma(\Omega)= ({\bf A}\Omega+{\bf B})({\bf C}\Omega+{\bf D})^{-1}\ .
\ee 
A paramodular form ${\Phi}(\Omega)$ of weight $k$ is a holomorphic function on the Siegel upper half plane that satisfies
\be\label{eq:tz}
{\Phi}( ({\bf A}\Omega+{\bf B})({\bf C}\Omega+{\bf D})^{-1})=\det({\bf C}\Omega+{\bf D})^k {\Phi}(\Omega)~.
\ee 
We denote by $M_k(\Gamma_t^+)$ the space of Siegel paramodular
forms of weight $k$ under $\Gamma_t^+$. 
Note that $t=1$ corresponds to $\Gamma_1^+=Sp(4,\Z)$, which is the more familiar Siegel modular group. 

It is useful to characterize some elements $\gamma$ explicitly.
First we note that
\be\label{eq:flip}
\gamma=\left(
\begin{array}{cccc}
 0 & \sqrt{t} & 0 & 0 \\
 {1\over \sqrt{t}} & 0 & 0 & 0 \\
 0 & 0 & 0 & {1\over \sqrt{t}} \\
 0 & 0 & {\sqrt{t}} & 0 \\
\end{array}
\right)~,
\ee
exchanges $\tau$ and  $t \rho$ as in \eqref{eq:s11}. $\Gamma_t^+$ contains also $SL(2,\ZZ)$ as a subgroup; the element is given by
\be\label{SLelement}
\gamma=\left(
\begin{array}{cccc}
 a & 0 & b & 0 \\
 0 & 1 & 0 & 0 \\
 c & 0 & d & 0 \\
 0 & 0 & 0 & 1 \\
\end{array}
\right)~,\quad {\rm with }~~ad-bc =1~,
\ee
which gives the coordinate transformation
\be
\tau \mapsto \frac{a\tau+b}{c\tau+d}~,\qquad
z \mapsto \frac{z}{c \tau +d}~, \qquad
\rho \mapsto \rho -\frac{c z^2}{c \tau +d } ~.
\ee
This gives the correct transformation behavior
for $\varphi_{k,m}$ in \eqref{eq:jf1}. Moreover the transformation
\be\label{shiftelement}
\gamma=\left(
\begin{array}{cccc}
 1 & 0 & 0 & \mu  \\
 \lambda  & 1 & \mu  & 0 \\
 0 & 0 & 1 & -\lambda  \\
 0 & 0 & 0 & 1 \\
\end{array}
\right)~,
\ee
leads to the other transformation property for Jacobi forms in \eqref{eq:jf2}.
The $\varphi_{k,m}$ are then indeed Jacobi forms.
Finally note that $\Phi \in M_k(\Gamma_t^+)$ has to be invariant under
\be
\left(\begin{array}{cccc} 
	1&0&0&0\\
	0&1&0&t^{-1}\\
	0&0&1&0\\
	0&0&0&1
\end{array}\right)~,
\ee
which means that all non-vanishing powers of $p$ 
are multiples of $t$. It follows from \eqref{eq:s10} that $\Phi$
gives a family of Jacobi forms with index $t\ZZ^+$.%

Before moving on, it is worth mentioning a few properties and nomenclature that expand the notion of Jacobi forms in \eqref{eq:jf1}-\eqref{eq:jf2} to modular objects that are not neccessarly holomorphic. First, we define the \emph{discriminant} $\Delta:= 4nm-l^2$.
The coefficients $c(n,l)$ in \rref{eq:vp} only depend on $\Delta$ and $l$ (mod $2m$), and in fact only on $\Delta$ if $m$ is prime. A polar state in $\varphi_{k,m}(\tau,z)$ is one with $\Delta<0$.

We will denote the space of Jacobi forms of weight $k$ and index $m$ by $J_{k,m}$.  
There are several special cases and generalizations 
of Jacobi forms which have to do with the summation range
in (\ref{eq:vp}). \emph{Jacobi cusp forms} are Jacobi forms
for which $c(0,l)=0$. In particular they vanish at the cusp
$\tau = i\infty$.  \emph{Weak Jacobi forms} are holomorphic functions that
satisfy (\ref{eq:jf1}) and (\ref{eq:jf2}), and we have $c(n,l)=0$ if
$\Delta < - m^2$. \emph{Nearly holomorphic Jacobi forms}  are allowed to have a
pole at the cusp $q=0$. 
In total we thus have the inclusions
\be
J_{k,m}^{cusp} \subset J_{k,m} \subset J_{k,m}^{weak} \subset J_{k,m}^{nh}\ .
\ee

So far we have defined Siegel paramodular forms to be holomorphic on the Siegel upper half plane which implies that the associated $\varphi_{k,m}$ are Jacobi forms. For our applications this is too restrictive, as we want exponential growth of the coefficients. For this reason, we will allow for meromorphic forms. The associated $\varphi_{k,m}$ are then no longer Jacobi forms, but rather weak Jacobi forms or even meromorphic Jacobi forms. We will however only consider very particular cases of meromorphic paramodular forms, as we will discuss now.

\subsection{Exponential lifts}\label{sec:explift}

An exponential lift is a simple way to build a paramodular form starting from a Jacobi form.  As we will see later, it has a natural interpretation as a symmetric orbifold of a CFT$_2$ that can lead to a holographic interpretation.

The \emph{exponential lift} is described in Theorem 2.1 
of \cite{MR1616929}, which states:
\begin{quote}
Let $\varphi\in J^{nh}_{0,t}$ be a nearly holomorphic Jacobi form of weight 0 and index $t$
with integral coefficients
\be
\varphi(\tau,z)= \sum_{n,l} c(n,l)q^n y^l\ .
\ee
Define
\be\label{eq:abc}
A = \frac{1}{24}\sum_l c(0,l)~,\qquad  B = \frac{1}{2}\sum_{l>0} l c(0,l)~, \qquad C = \frac{1}{4}\sum_l l^2 c(0,l)\ .
\ee
and 
\be\label{eq:k1}
k={1\over 2} c(0,0)~.
\ee
Then the exponential lift of $\varphi$ is the product
\be\label{explift}
\textrm{Exp-Lift}(\varphi)(\Omega)= q^A y^B p^C \prod_{\substack{n,l,r\in\ZZ
		\\(n,l,r)>0}} (1-q^n y^l p^{tr})^{c(nr,l)}\ ,
\ee
where $(n,l,r)>0$ means $r >0 \lor (r=0 \land n>0) \lor (n=r=0 \land l <0)$, and it defines
a meromorphic modular form of weight $k$ with respect to $\Gamma^+_t$.
It has a character (or a multiplier system if the weight is half-integral) induced by $v^{24A}_\eta \times v^{2B}_H$.
Here $v_\eta$ is a 24th root of unity, and $v_H=\pm1$.
\end{quote}
Even though we stated the theorem for nearly holomorphic forms, we will only use it for weak Jacobi forms, in which case we actually have $C=tA$.

The exponential lift can be naturally split into two
factors, namely
\be\label{explift2}
\textrm{Exp-Lift}(\varphi)(\Omega)= q^A y^B p^C 
\prod_{\substack{(n,l)>0} }(1-q^n y^l)^{c(0,l)}\times
\prod_{\substack{n,l,r\in\ZZ
		\\r>0}} (1-q^n y^l p^{tr})^{c(nr,l)}\ .
\ee
Here $(n,l)>0$ means $n>0 \lor (n=0\land l<0)$. The first factor is usually denoted as the Hodge factor, and it is defined as
\be
\phi_{k, C}(\tau,z)=q^{A} y^B \prod_{\substack{(n,l)>0} }(1-q^n y^l)^{c(0,l)}~,
\ee
with weight $k$ as given in \eqref{eq:k1}  and index $C$ in \eqref{eq:abc}. The inverse of $\phi_{k,C}$ is Jacobi-like form with a multiplier system.
The second factor can be naturally written in terms
of Hecke operators $T_-(r)$, namely as
\be\label{eq:symprod}
\prod_{\substack{n,l,r\in\ZZ
		\\r>0}} (1-q^n y^l p^{tr})^{c(nr,l)} =\exp\left( - \sum_{r\geq1} r^{-1}p^{tr}\varphi|T_-(r) \right) \equiv \frac{1}{Z^\varphi_{Sym}}\ .
\ee
If $\varphi$ is some elliptic genus or
partition function $\chi$ of a CFT,
then $Z^\chi_{Sym}$ is the generating function
for the partition functions of the symmetric orbifolds
of that theory:
\be\label{explift3}
Z^\chi_{Sym}= \sum_{r=0}^\infty p^{tr} \chi(\tau,z; {\rm Sym}^r(M))=\frac{p^C\phi_{k, C}(\tau,z)}{\textrm{Exp-Lift}(\chi)(\Omega)}
\ee
Note that all powers of $p$ are multiples of $t$. If $\chi(\tau,z; M)$ is a weak Jacobi of index $t$, then $\chi(\tau,z; {\rm Sym}^r(M))$ has index $tr$.

\subsection{Zeros and poles}\label{sec:zp}

The most important component of our analysis in subsequent sections relies on the divisors of the paramodular forms constructed via \eqref{explift}. This is the second portion of Theorem 2.1  in \cite{MR1616929}, which we now summarize.

For paramodular forms that have a product expansion, such as \eqref{explift}, it is rather simple to identify some of the divisors: choosing $\tau, z, \rho$ such that $q^n y^l p^{tr}=1$ in one of the factors will make that factor vanish, so that the product either vanishes or diverges. Because of the invariance under $\Gamma_t^+$, 
divisors will always come as orbits of $\Gamma_t^+$, and are known as \emph{Humbert surfaces}. 
These surfaces, denoted as
$H_D(b)$, can always be written as 
\be\label{eq:hdb}
H_D(b) = \pi^+_t(\{\Omega \in \H_2: a\tau + bz + t\rho =0 \} )\ ,
\ee
where $\pi^+_t$ is the set of images of $\Gamma^+_t$.
Here $a,b \in \ZZ$, the discriminant $D$ is given by $D= b^2- 4ta$ and $b$ is defined $\mod 2t $.
Each such divisor comes with multiplicity (or degree) $m_{D,b}$. The total divisor of the exponential
lift (\ref{explift}) is given by the Humbert surfaces
\be
\sum_{D,b} m_{D,b} H_D(b)~,
\ee
where the multiplicities $m_{D,b}$ are given by
\be
m_{D,b}= \sum_{n>0} c(n^2a,nb)\ ,
\ee
where $c(n,l)$ are the Fourier coefficients of
the underlying form $\varphi$.
From this we see that the Humbert surface of maximal
discriminant $D$ comes from the term with maximal
polarity of $\varphi\in J^{nh}_{0,t}$.

Following Sec.\ 1.3 of \cite{MR1616929}, it is also useful to note that for
 $\ell = (e,a,-\frac{b}{2t},c,f)$ with
$e,a,b,c,f \in \Z$ and $(e,a,b,c,f)=1$, and the discriminant 
\be\label{eq:DD}
D(\ell)=2t(\ell,\ell) = b^2 -4tef -4tac~,
\ee
there is a natural action of $\Gamma_t^+$ on $\ell$
that leaves $D(\ell)$ invariant. $\ell$ then
defines a divisor in $\H_2$ via the quadratic equation
\be\label{Humzero}
tf (z^2-\tau\rho) + tc\rho +bz+a\tau+e =0\ .
\ee
These are the most general images in \eqref{eq:hdb}.

\section{Five special examples}\label{sec:five}

Let us now turn to specific examples of paramodular forms.
We are interested in forms that have the potential of providing a novel setup for holography.   
We will focus on paramodular forms that are built from exponential lifts: those always have an extended Cardy regime which we can associate to black hole growth, as they are  connected to symmetric product orbifolds \cite{Keller:2011xi}. Within such exponential lifts, there is a class whose poles are parametrized by  the Humbert surface $H_1(1)$.
This class has two appealing features: the formula for the degeneracy of black hole states admits a particularly simple and elegant formula, and the Cardy regime benefits from an even larger extension than a generic symmetric product \cite{Belin:2016knb}. In this section we will carefully characterize these forms, with the aim in later sections to extract properties of these counting formulas that are relevant to make a comparisson with the putative supergravity regime.  

 Let us emphasize however that restricting to $H_1(1)$ is neither fundamental nor exhaustive. In fact the examples in  this section will illustrate that despite their specificity, there is plenty of room for interesting constructions.

In the following we will present five explicit examples of paramodular forms whose \emph{only} divisor is $H_1(1)$. Note that they will be exponential lifts of strictly weak Jacobi forms.\footnote{ We note that for a nearly holomorphic Jacobi form the negative powers of $q$ will generate poles that have $a\neq0$ in \eqref{eq:hdb}.} To describe these examples, it is convenient to introduce some notation. We will study the growth of the Fourier coefficients in 
\be\label{eq:phikrec}
{1\over {\Phi_k}(\Omega)} = \textrm{Exp-Lift}(-\varphi)(\Omega)~,
\ee
where the minus sign (and reciprocal) is mostly conventional: it is common to define the first few coefficients of $\varphi$ as positive and $\Phi_k(\Omega)$ is thought as a cusp form. As before, $k$ is the weight of the paramodular form $\Phi_k(\Omega)$. The coefficients of the seed $\varphi \in J^{weak}_{0,t}$ are defined via
\be
\varphi_{0,t}(\tau,z)=\sum_{n,l} c(n,l) q^n y^l~, 
\ee
and as for any weak Jacobi form, we have the `Witten index' identity
\be\label{eq:wittenid}
\sum_l c(n,l) = 0\,,\quad \forall\, n>0~,
\ee
which later on will allow us to read off rather easily the residues dictated by $H_1(1)$.
The coefficients of the paramodular form will be parametrized as
\be\label{eq:s113}
{1\over {\Phi_k}(\Omega)} =\sum_{m,n,l} c(m,n,l) p^m q^n y^l~. 
\ee

From the discussion in Sec.\ \ref{sec:zp}, the Humbert surface $H_1(1)$ is given by
\be
H_1(1) = \pi^+_t(\{\Omega \in \H_2:  z + t\rho =0 \} )\ ,
\ee
with 
\be\label{eq:m11}
m_{1,1}={\sum_{l<0} c(0,l)}~.
\ee
For any $H_1(1)$ surface, we will have a divisor at $z=0$, since the transformation 
\be
\left(
\begin{array}{cccc}
 1 & t & 0 & 0 \\
 0 & 1 & 0 & 0 \\
 0 & 0 & 1 & 0 \\
 0 & 0 & -t & 1 \\
\end{array}
\right) \in \Gamma_t^+\ ,
\ee
maps $z\mapsto z+t\rho$. The behavior near $z=0$ will be vital as we extract the Fourier coefficients; the leading behavior around $z=0$, up to numerical coefficients, is 
\begin{align}\label{liftresidue}
{\Phi_k}(\Omega)\,& \sim\, q^A p^{tA} \prod_{r>0} (1- p^{tr})^{24A} \prod_{n>0} (1-q^n)^{24A} \prod_{l<0} (1-y^l)^{c(0,l)}\cr
&\sim z^{m_{1,1}} q^A p^{tA} \prod_{m>r} (1- p^{tr})^{24A} \prod_{n>0} (1-q^n)^{24A}\cr
&= z^{m_{1,1}} \eta(\tau)^{24A} \eta(t\rho)^{24A}
\end{align}
where we used the identity \eqref{eq:wittenid}, and $\eta(\tau)$ is the Dedekind-eta function.

The restriction of having only one type of Humbert surface puts tight constraints on the seed $\varphi_{k,m}$. In particular, its  coefficients must satisfy
\be\label{eq:rest1}
c(0,l) =0 \,,\quad \forall\, |l|>1~,
\ee
which assures that $m_{b^2,b}=0$ for $b\neq 1$. To have a pole rather than a zero at the divisor, we need $c(0,1)>0$. Together with \eqref{eq:wittenid}, this means that the parameters in \eqref{eq:abc} and \eqref{eq:m11} simplify to
\be\label{eq:abc1}
A={1\over 24} \le(2c(0,1)+c(0,0)\ri)~, \quad C=B= {1\over 2} c(0,1)~,  \quad m_{1,1}=c(0,1)~,
\ee
where we used $c(0,1)=c(0,-1)$. Because $C=tA$, we have
\be
c(0,0)= \le({12\over t} -2\ri)c(0,1)~.
\ee
For reasons that we will explain in the Sec.\ \ref{sec:method}, we want to further focus on 
\be\label{eq:m112}
m_{1,1}=2~,
\ee
which due to \eqref{eq:abc1} gives $C=B=1$ and $A=1/t$. This generates a second order zero in $\Phi_k$ and hence a double pole around $z=0$ in the reciprocal \eqref{eq:phikrec}. In addition,  the weight of $\Phi_k(\Omega)$ is given by
\be\label{eq:k2}
k={12\over t} -2~,
\ee
where we used \eqref{eq:k1}. Provided we want only positive and integral weights, we have a tight range for the index: $$t=1,2,3,4,6~.$$ These are our \emph{five special examples}, which we describe individually in the following.

All the mathematical properties of these examples were collected from \cite{Gritsenko:1999fk}, and below we display the main highlights adjusted to our discussion. The usual interpretation of these forms is as the elliptic genus of a non-linear sigma model with target space a Calabi-Yau manifold in $d$-complex-dimensions (CY$_d$). Without the restriction in \eqref{eq:k2}, the elliptic genus of CY$_d$ is a linear combination of the finite dimensional space of weak Jacobi form of weight zero and fixed index. The linear combination is dictated by the topological data of CY$_d$; our condition \eqref{eq:k2} can be interpreted as a restrictions on the hodge numbers of the CY$_d$.

\paragraph{t=1, k=10.}  The most well known example that we will study is  the exponential lift of 
\begin{align}\label{eq:p01}
\phi_{0,1}&=4\le({\theta_2(\tau,z)^2\over \theta_2(\tau)^2}+{\theta_3(\tau,z)^2\over \theta_3(\tau)^2}+{\theta_4(\tau,z)^2\over \theta_4(\tau)^2}\ri) \cr&= y^{-1}+10+y+(10y^{-2} -64y^{-1}+108-64y+10y^2)q  +\ldots ~,
\end{align}
which is the unique weak Jacobi form of weight zero and index one. Here $\theta_i(\tau,z)$ are the usual theta functions, and $\theta_i(\tau)\equiv\theta_i(\tau,0)$.  The 
resulting paramodular form is 
\begin{align}\label{eq:t1k10}
{1\over \Phi_{10}}= \textrm{Exp-Lift}(-2\phi_{0,1})(\Omega)~,
\end{align}
where $\Phi_{10}$ is the famous Igusa cusp form. (We recall that for $t=1$ the paramodular group is simply $Sp(4,\ZZ)$.) The seed, $2\phi_{0,1}$, is the elliptic genus of a $K3$ surface. The left hand side of \eqref{eq:t1k10} is known as the DVV formula \cite{Dijkgraaf:1996it}, which counts 1/4 BPS states in ${\cal N}=4$ string theory in four dimensions. Generalizations of this formula to CHL models are discussed in \cite{JatkarSen2006a,DavidSen2006,PaquetteVolpatoZimet2017}, which corresponds to orbifolds of $K3$.   

\paragraph{t=2, k=4.} The next example comes from studying the exponential lift of
\begin{align}
\phi_{0,2}&={1\over 2} \eta(\tau)^{-4} \sum_{m,n\in \ZZ} (3m-n)\le(-4\over m\ri)\le(12\over n\ri)q^{(3m^2+n^2)/24}y^{(m+n)/2}\cr
&=y^{-1}+4+y+(y^{-3}+8y^{-2} -y^{-1}+16-y+8y^2+y^3)q+\ldots~, 
\end{align}
whose paramodular form is
\begin{align}
{1\over \Phi_4}= \textrm{Exp-Lift}(-2\phi_{0,2})(\Omega)~.
\end{align}

The weak Jacobi form $\phi_{0,2}$ appears in the construction of the elliptic genus of CY$_4$, and it is interesting to display some properties of this lift. For an arbitrary CY$_4$, the second quantized elliptic genus is
\be\label{eq:expcy4}
\textrm{Exp-Lift}(-\chi_0(M) \psi_{0,2})\times \textrm{Exp-Lift}(\chi_1(M) \phi_{0,2})~,
\ee
where, for a K\"{a}hler manifold $M$, we have introduced the topological character
\be\label{eq:defchim}
\chi_r(M)\equiv\sum_s(-1)^{s} h^{r,s}(M)~,
\ee
which is a combination of the Hodge numbers $h^{r,s}$ of $M$. The second weak Jacobi form in \eqref{eq:expcy4} is given by
\be
\psi_{0,2}= \phi_{0,1}^2-20\,\phi_{0,2}~.
\ee
The divisor of  \eqref{eq:expcy4} is equal to
\be
(\chi_1(M)-\chi_0(M))H_1(1) -\chi_0(M) H_4(2)~.
\ee 
The interesting aspect of this divisor, is that it is rather simple to generalize our conditions to design an example that has  poles described by $H_1(1)$, and zeroes dictated by $H_4(2)$: it would simply require that $\chi_0(M)<0$ and $\chi_0(M)-\chi_1(M)>0$. Our focus here will be limited to $m_{1,1}=2$ and $k>0$, which restricts $\chi_0(M)=0$ and $\chi_1(M)=-2$. Note that we could generalize our analysis by keeping the pole given by $H_1(1)$, but allowing zeroes as well. For example, one could consider $\chi_0(M)=-1$ and $\chi_1(M)=-3$. We expect the presence of zeroes to change some aspects of our results and we will comment on it in Sec. \ref{sec:discussion}. 

\paragraph{t=3, k=2.} For our next example, the unique weak Jacobi form is 
\begin{align}
\phi_{0,3}&= \phi_{0,{3\over 2}}^2(\tau,z)\cr
&=y^{-1}+2+y-(2y^{-3}+2y^{-2} -2y^{-1}+4-2y+2y^2+2y^3)q+\ldots~, 
\end{align}
where
\begin{align}\label{eq:p03/2}
\phi_{0,\frac{3}{2}}(\tau,z)&=\frac{\theta(\tau,2z)}{\theta(\tau,z)}\cr
&=y^{-1/2}\prod_{n=1}^\infty(1+q^{n-1}y)(1+q^n y^{-1})(1-q^{2n-1}y^2)(1-q^{2n-1}y^{-2})~,
\end{align}
and 
\be\label{eq:theta}
\theta(\tau,z)=-q^{1/8}y^{-1/2}\prod_{n=1}^\infty(1-q^{n-1}y)(1-q^n y^{-1})(1-q^{n})~.
\ee
The resulting lift is
\begin{align}
{1\over \Phi_2}= \textrm{Exp-Lift}(-2\phi_{0,3})(\Omega)~.
\end{align}
We emphasize that the only divisor of this paramodular form is $H_1(1)$.

The index $t=3$ form $\phi_{0,3}$ is closely related to the elliptic genus of CY$_6$. More explicitly, it is a combination of three weak Jacobi forms, which are
\begin{align}
&\phi_{0,3}~, \cr 
 &\psi_{0,3}^{(2)}= \phi_{0,1}\phi_{0,2}-15\, \phi_{0,3}~,\cr 
 &\psi_{0,3}^{(3)}= \phi_{0,1}^3-30\,\phi_{0,1}\phi_{0,2}+117\,\phi_{0,3}~,
\end{align}
and the second quantized elliptic genus of CY$_6$ reads
\be\label{eq:expcy6}
\textrm{Exp-Lift}(-\chi_0(M) \psi^{(3)}_{0,3})\times \textrm{Exp-Lift}(\chi_1(M) \psi^{(2)}_{0,3})\times  \textrm{Exp-Lift}((\chi_1(M)-\chi_2(M)) \phi_{0,1})~.
\ee
 The characters $\chi_r(M)$ are defined as in \eqref{eq:defchim}. The divisor characterizing this exponential lift is
 \be
(\chi_1(M)-\chi_2(M)-\chi_0(M)) H_1(1) + \chi_1(M) H_4(2) -\chi_0(M) H_9(3)~. 
 \ee
Our case of interest, sets $\chi_1(M)=\chi_0(M)=0$ and $\chi_2(M)=2$. However, as we advocated before, generalizations that accommodate for $\chi_{0,1}(M)\neq0$ while keeping $H_1(1)$ as the only source of poles would be interesting in future studies.

\paragraph{t=4, k=1.} There is one \emph{odd} example in our construction, which is given by the exponential lift of the weak Jacobi form
\begin{align}
\phi_{0,4}&={\theta(\tau,3z)\over \theta(\tau,z)}\cr &=y^{-1}+1+y-(y^{-4}+y^{-3}-y^{-1}-2-y+y^3+y^4)q+\ldots~,
\end{align}
with $\theta(\tau,z)$ as defined in \eqref{eq:theta}. The resulting paramodular form is
\begin{align}
{1\over \Phi_1}= \textrm{Exp-Lift}(-2\phi_{0,4})(\Omega)~,
\end{align}
the only paramodular form of odd weight in our list.\footnote{Note that the multiplier system of  $\Phi_1$ is non-trivial according to the theorem in Sec. \ref{sec:explift}.} It is interesting to note that $\Phi_1$, in addition to the product expansion in \eqref{explift}, has a simple expansion as 
\be
\textrm{Exp-Lift}(\phi_{0,4}) = {1\over 2} \sum_{n,m\in \ZZ} \le({-4\over n}\ri) \le({-4\over m}\ri) q^{n^2/8} y^{nm/4} s^{m^2/8}~.
\ee

$\phi_{0,4}$ is one of four basic weak Jacobi forms of index 4 that characterizes the elliptic genus of CY$_8$. The most general second quantized elliptic genus is rather involved (but straightforward), and we refer to \cite{Gritsenko:1999fk} for explicit expressions. 

\paragraph{t=6, k=0.} Our last example is given characterized by the weak Jacobi form
\begin{align}
\phi_{0,6}&=\phi_{0,2}\phi_{0,4}-\phi_{0,3}^2\cr
&=y^{-1}+y+(-y^{-5}+y^{-1}+y-y^5)q +\ldots
\end{align}
We note that $\phi_{0,6}(\tau,z)=\phi_{0,{3\over 2}}(\tau, 2z)$ as defined in \eqref{eq:p03/2}. We denote the exponential lift as
\begin{align}\label{eq:t6cy3}
{1\over \Phi_0}= \textrm{Exp-Lift}(-2\phi_{0,6})(\Omega)~.
\end{align}
It is interesting to note that for $t=6$ we cannot impose that the \emph{only} divisor is $H_1(1)$: the lift of  $\phi_{0,6}$ has  
\be
H_1(1) - H_1(5)~,
\ee
and hence ${1/ \Phi_0}$ has non-trivial zeroes in addition to the poles dictated by $H_1(1)$. 

The most naive interpretation of this form is of course as coming from the elliptic genus of a CY$_{12}$. We can however also interpret it as coming from a CY$_{3}$: the elliptic genus of CY$_3$ is given by
\be
{1\over 2}e(CY_3)\phi_{0,{3\over 2}}(\tau, z)~,
\ee
where $e(CY_3)$ is the Euler number of the manifold. By rescaling $z\mapsto 2z$ this then becomes $\phi_{0,6}$.
In contrast to \eqref{eq:p01}--\eqref{eq:t1k10}, note that $\Phi_0$  does not count BPS states in the MSW string \cite{Maldacena:1997de} that are relevant to 4D BPS black holes in ${\cal N}=2$ supergravity, nor M-theory backgrounds of the form AdS$_3\times S^2 \times CY_3$.\footnote{There are at least two pieces of evidence to claim we are not describing such black holes: the logarithmic corrections in \cite{Sen2012b} do not match those predicted by \eqref{eq:t6cy3} as we showed in \cite{Belin:2016knb}; the form \eqref{eq:t6cy3} does not seem to capture the attractor flows in \cite{DenefMoore2011}. We will elaborate more about this in Sec. \ref{sec:sugra}. }

It is interesting that our restriction to exponential lifts with a double pole at $H_1(1)$ as the only divisor gives finite number of examples. These are the forms which we will study in the subsequent sections, and in table \ref{table:thefive} we list their basic data.

\noindent\begin{minipage}{\textwidth}
	\begin{center}
		\begin{tabular}{|c|c|c|c|c|c|c|}
		\hline
		 &Seed ($\varphi$) & Weight ($k$) & Group & $A$ & $B$ & $C$ \cr 
			\hline\hline
	 $\Phi_{10}$	&$2\phi_{0,1}$ & 10 & $SP(4,\ZZ)$& 1 &1 &1\cr
 			\hline
	 $\Phi_4$	&$2\phi_{0,2}$ & 4 & $\Gamma_2^+$& 1/2 & 1&1\cr	
			\hline
	 $\Phi_2$	&$2\phi_{0,3}$ & 2 & $\Gamma_3^+$& 1/3& 1&1\cr
	 		\hline
	 $\Phi_1$	&$2\phi_{0,4}$ & 1 & $\Gamma_4^+$& 1/4& 1&1\cr
	 \hline
	  $\Phi_0$	&$2\phi_{0,6}$ & 0 & $\Gamma_6^+$& 1/6 & 1&1\cr
	  			\hline
					\end{tabular}\captionsetup{justification=raggedright}\captionof{table}{The five exponential lifts whose only divisor is $H_1(1)$, and integral weight that is non-negative. The coefficients $A,B,C$  and $k$ are those define in \eqref{eq:abc} and \eqref{eq:k1}. 
					}\label{table:thefive}
	\end{center}
	\vspace{.2em}
\end{minipage}

%%%%%%%%%%%%%%%%%%%%%%%%%%%%%%%%%%%%%%%%%%%%%%%%%%%%%%%%%%%%%%%%%%
%%%%%%%%%%%%%%%%%%%%%%%%%%%%%%%%%%%%%%%%%%%%%%%%%%%%%%%%%%%%%%%%%%
%%%%%%%%%%%%%%%%%%%%%%%%%%%%%%%%%%%%%%%%%%%%%%%%%%%%%%%%%%%%%%%%%%

\section{Methodology}\label{sec:method}

In this section we present our methodology to extract the Fourier coefficients for the five special examples listed in Table \ref{table:thefive}. Following the notation in Sec. \ref{sec:five}, we are interested in obtaining the Fourier coefficents $c(m,n,l)$ defined via\footnote{Note that following \rref{explift}, the powers of $p$ are given by $tk-1, \ k\in\mathbb{N}$.}
\be\label{eq:s111}
{1\over {\Phi_k}(\Omega)} =\sum_{m,n,l} c(m,n,l) p^m q^n y^l~. 
\ee
It is useful to define the discriminant $\Delta$ of a state,
\be
\Delta \equiv 4nm-l^2~,
\ee	
which is invariant under the action of \eqref{shiftelement}. We then have states with
\be
\Delta \geq 0 ~: \quad {\rm positive~or~zero~ discriminant ~state}~.
\ee
Extracting $c(m,n,l)$ for states with $\Delta\gg 1$ is the focus of \cite{Belin:2016knb}. Here instead we will focus on states with 
\be
\Delta <0 ~:\quad {\rm negative~ discriminant~state}~.
\ee

The techniques and features used to evaluate $c(m,n,l)$ are sharply different for positive versus negative states.\footnote{Appendix \ref{app:bh} has a brief review of the contour used for states with $\Delta\gg1$.} In the following we review the method that was used in \cite{Sen:2011mh} to obtain the degeneracy of negative discriminant states for $t=1$, and generalize it to higher values of $t$ that are relevant for our five examples. 

\subsection{Tessellation of $\H_2$ by $\Gamma_t^+$}

Fundamentally we want to obtain $c(m,n,l)$ from a contour integral such as 
\begin{align} \label{cdef}
c(m,n,l)=\oint_{p=0} {dp\over 2\pi i p} \oint_{q=0} {dq\over 2\pi i q} \oint_{y=0} {dy\over 2\pi i y} \, \frac{1}{\Phi_k(\Omega)} p^{-m} q^{-n} y^{-l} \,.
\end{align}
However, we need to be careful about our choice of contour here. Since $1/\Phi_k(\Omega)$ is meromorphic, $c(m,n,l)$ defined in this way depends on the precise choice of the contour \cite{Cheng:2007ch}. 
A very simple illustration of this ambiguity can be described as follows. For our five examples there is a pole at $z=0$. To get the Fourier coefficient, we have to expand
\be\label{eq:yexp}
(1-y^l)^{-c(0,l)}~, \qquad c(0,l)>0~.
\ee
If $l \,\text{Im}z  >0$,  then the correct expansion of \eqref{eq:yexp} is some geometric series in powers of $y$. If however $l\,\text{Im} z <0$ then we need to write \eqref{eq:yexp} as
\be\label{eq:yexp1}
 (-1)^{-c(0,l)} y^{-lc(0,l)}(1-y^{-l})^{-c(0,l)}~, \qquad c(0,l)>0~,
\ee
which gives an expansion in powers of $y^{-1}$ instead. This ambiguity is what forces us to define chambers, within which the expansion is convergent and the coefficients do not suffer from ambiguities do to the crossing of a pole in the function.

To start, $\Phi_k$ is defined on the Siegel upper half plane  $\H_2$, defined in \eqref{eq:uhp}, which  implies 
\be\label{eq:uhp1}
\text{Im} \tau >0 \,, \qquad \text{Im} \rho >0 \,, \qquad \text{Im} \tau\text{Im} \rho - (\text{Im} z)^2>0 \,.
\ee
We can deform our contours on this domain without changing (\ref{cdef}) as long as we do not cross any poles when doing so. We therefore want to tessellate $\H_2$ into chambers whose boundaries are given by the poles of $1/\Phi_k$. This will define regions (chambers) where we can accurately  evaluate \eqref{cdef}.

To define the chamber which will be useful for us, first note that
we want our contour to enclose $p=q=y=0$. For this choice we can take the real parts to be restricted to
\be\label{eq:re01}
0 \leq \text{Re} \tau, \text{Re} \rho, \text{Re} z <1 \,,
\ee
while for the imaginary parts we choose 
\be\label{eq:imlarge}
\text{Im} \rho~, \text{Im} \tau~, \text{Im} z \gg 0~.
\ee
Let us denote by  \textbf{R} the chamber which contains the point  $p=q=y=0$ and hence compatible with \eqref{eq:uhp1}--\eqref{eq:imlarge}. The boundaries of this region are defined as follows. As pointed out above, we need to pick an expansion around $y$ which is affected by the choice of sign of $l$; without loss of generality, we will choose 
\be\label{eq:imz0}
l<0~, \qquad \text{Im} z >0  \,.
\ee
 However we can still have some conflicts due to the images of $z=0$ under the paramodular group $\Gamma_t^+$; this will bound {\bf R} from below on the Siegel upper half plane.  
 The most general image of our pole is given by \eqref{Humzero}, and reads
  \be
 t\,f (z^2-\tau\rho) + t\,c\rho +bz+a\tau+e =0~,\qquad e,a,b,c,f \in \Z~.
 \ee
 Since we are interested in large imaginary values of the parameters as in \eqref{eq:imlarge}, we see that the only poles that we can encounter have $f=0$. This means we only have to consider linear poles. Therefore, our task is to
 characterize how the linear equation 
\be\label{eq:line1}
 t\,c\, \rho +b\,z+a\,\tau+e =0~,
\ee
 tessellates $\H_2$. More precisely, we can concentrate purely on the imaginary component of this equation,
 \be\label{eq:immaster}
  t\,c\, \text{Im} \rho +b\,\text{Im}z+a\,\text{Im}\tau =0~,
\ee
as the real component comes along for the ride.

\begin{figure}[htbp]
	\begin{center}
		\includegraphics[width=0.4\textwidth]{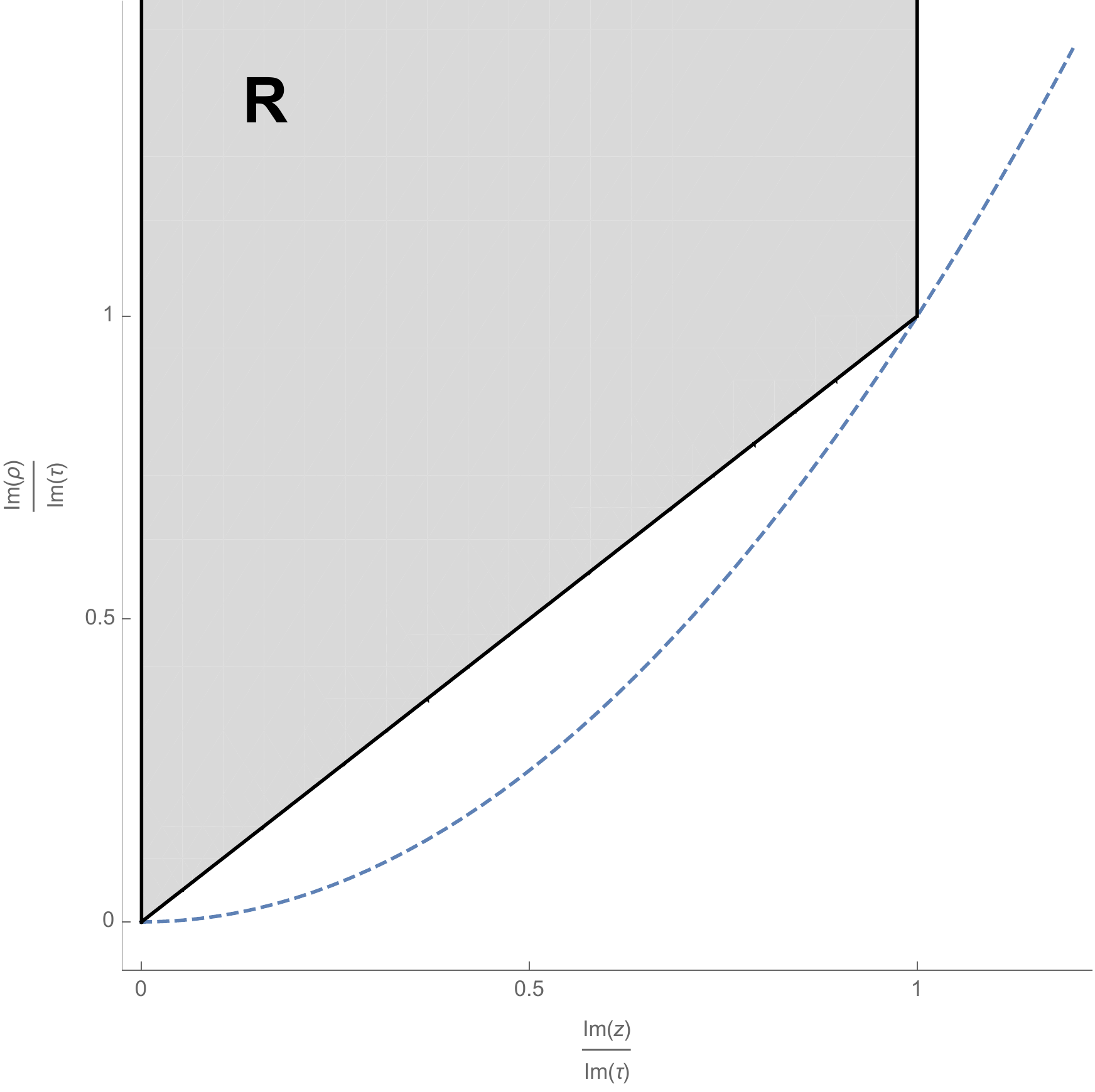}
		\includegraphics[width=0.4\textwidth]{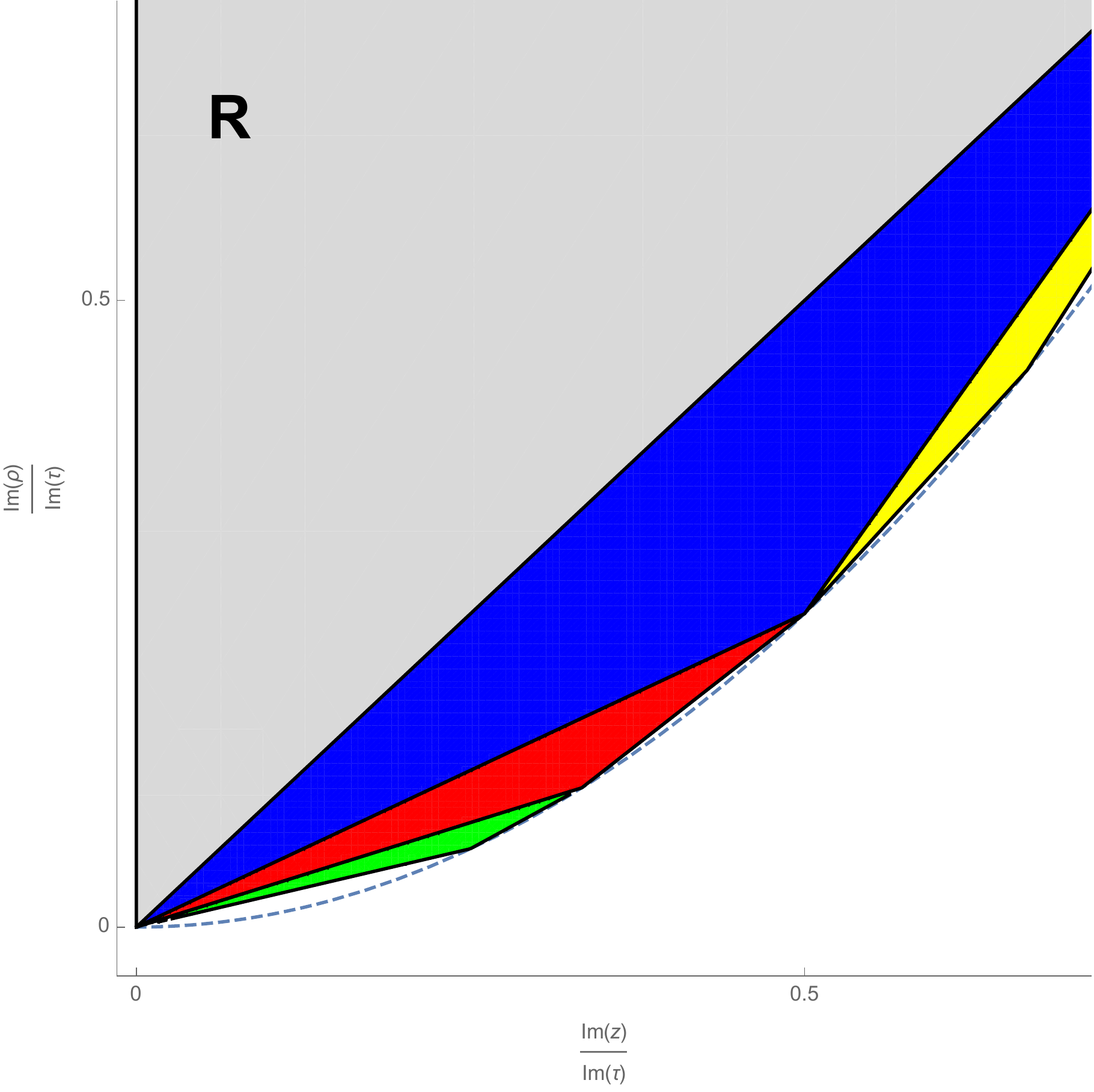}
		\caption{The plot to the left shows the chamber {\bf{R}} (shaded grey region) for $1/\Phi_{10}$, i.e. $t=1$. The dashed line corresponds to the boundary of the Siegel upper half plane; the solid lines correspond to \eqref{eq:lines0} . The plot to the right shows the tessellation of $\H_2$, which is produced by considering further lines in $H_1(1)$; each colored region represents a new chamber which is surrounded by the appropriate divisor in \eqref{eq:immaster}.}
		\label{chamberR}
	\end{center}
\end{figure}

\begin{figure}[htbp]
	\begin{center}
		\includegraphics[width=0.4\textwidth]{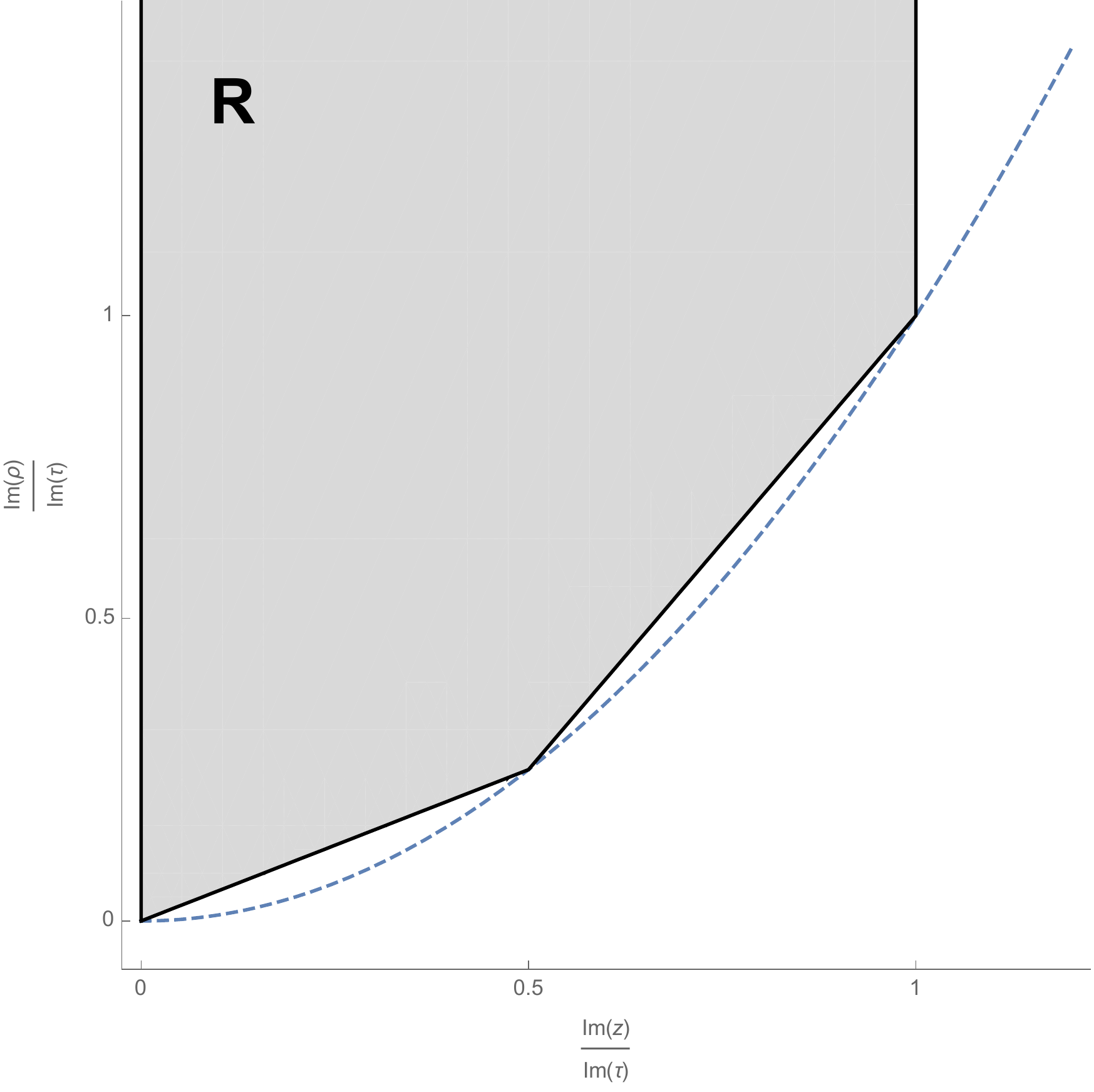}
		\includegraphics[width=0.4\textwidth]{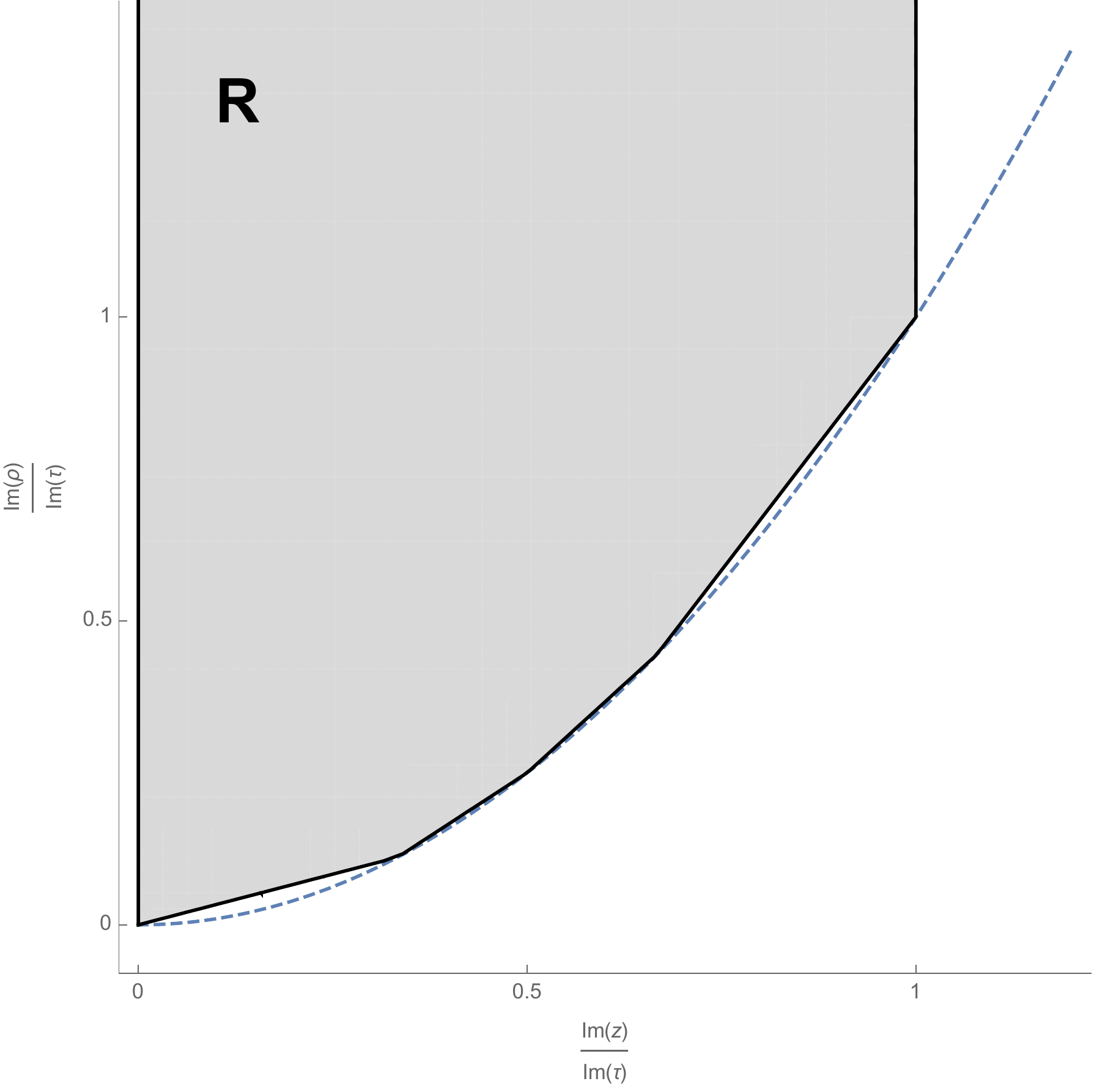}
		\caption{These plots show the chamber {\bf{R}} for $t=2$ (left) and $t=3$ (right). The dashed line corresponds to the boundary of the Siegel upper half plane; the region {\bf{R}} is always enclosed between the lines $\ima z=0$ and $\ima z=\ima \tau$ for any $t$. But the number of lines bounding {\bf{R}} from below depends on $t$: the relevant lines shown here are those in \eqref{eq:t2lines} and \eqref{eq:t3lines}.} 
		\label{teq2and3}
	\end{center}
\end{figure}

The lower boundary of {\bf R} will be dictated by the lines defined in \eqref{eq:line1}. To visualize the chambers in $\H_2$ it is useful to plot the Siegel upper half plane as a two-dimensional half plane in the ratios $\text{Im} \rho/\text{Im}\tau$ and $\text{Im} z/\text{Im} \tau$. 
Figure~\ref{chamberR} illustrates the shape of  \textbf{R} for $t=1$ and the tessellation below it;  Figure \ref{teq2and3} displays the analogous region {\bf R} for $t=2$ and $t=3$. 

\begin{figure}[htbp]
	\begin{center}
		\includegraphics[width=0.5\textwidth]{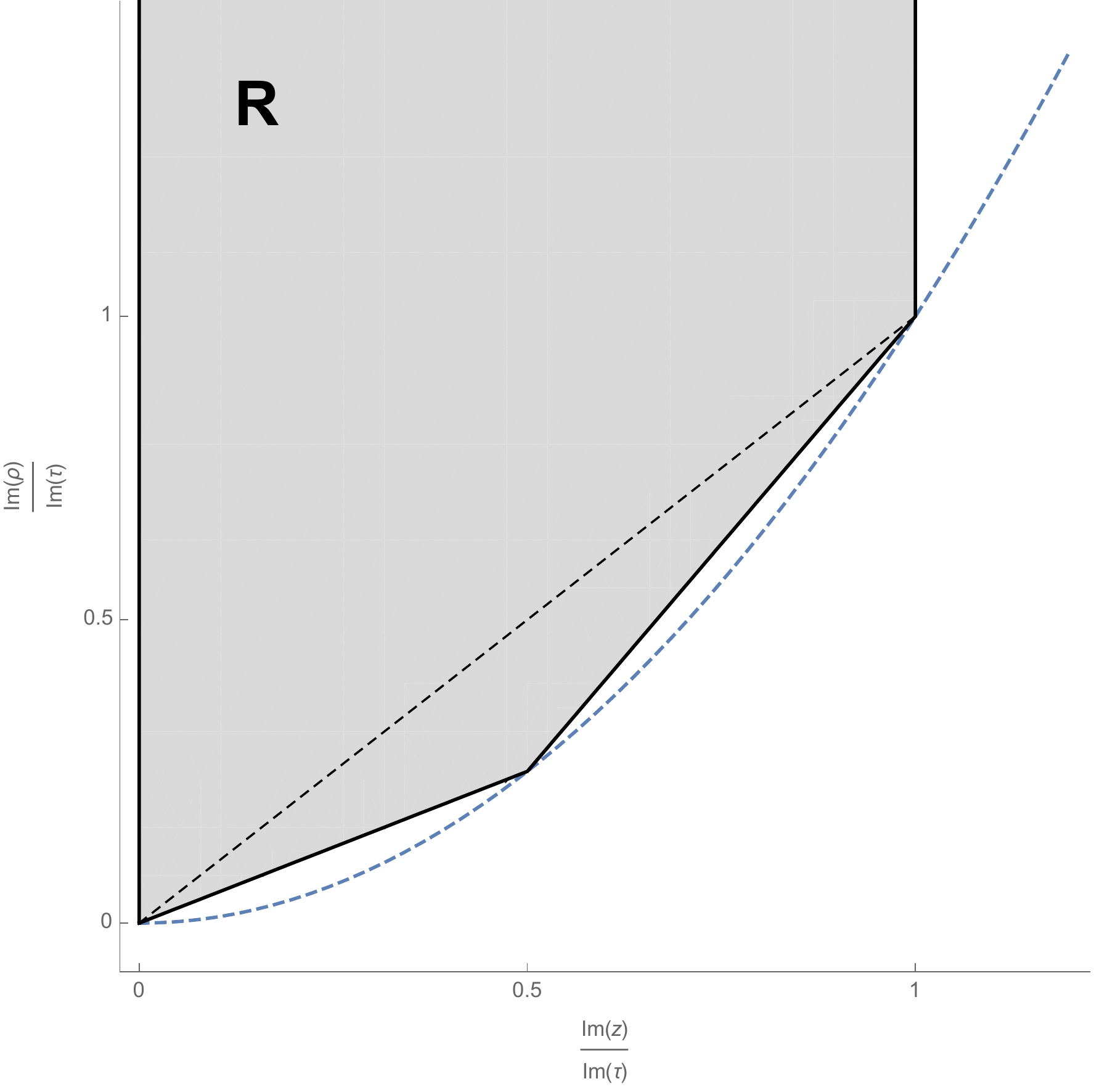}
		\caption{Here we show the chamber {\bf{R}} for the paramodular group $\Gamma_2^+$ in contrast to the analogous chamber for $SP(4,\ZZ)$. The blue dashed line is the boundary of the Siegel upper half plane. The black dashed line shows a linear pole for $t=1$ that is not a linear pole of the paramodular group $\Gamma_2^+$. The chamber {$\bf{R}$ }is the union of two chambers of $SP(4,\ZZ)$, lying to the left and right of the black dashed line. }
		\label{chamberRpara}
	\end{center}
\end{figure}

It is instructive to consider first the simplest example. For $t=1$ the poles that bound {\bf R} are 
\be\label{eq:lines0}
\text{Im}z=0 ~,\quad \text{Im}z=\text{Im}\rho~, \quad \text{Im}z=\text{Im}\tau~.
\ee
As plotted in Figure~\ref{chamberR}, the area enclosed by these three lines defines {\bf R} for $t=1$. For higher values of $t$, {\bf R} is always bounded by the two poles 
\be\label{eq:lines1}
\text{Im}z=0~,\quad \text{Im}z=\text{Im}\tau~.
\ee
The main difference for $t\neq1$ is however the number of segments that connect the origin ($\text{Im}z=\text{Im}\rho=0$) with the point $\text{Im}z=\text{Im}\tau=\text{Im}\rho$ (which is the intersection of the parabola in \eqref{eq:uhp1} with the second vertical line). For example, for $t=2$ we have the two poles
\bea\label{eq:t2lines}
 \ima z - 2\ima \rho &=& 0~, \cr
3 \ima z - 2\ima \rho - \ima \tau &=& 0~,  
\eea
while for $t=3$, there are four poles
\bea\label{eq:t3lines}
 \ima z - 3\ima \rho &=& 0~, \cr
5\ima z - 6\ima \rho - \ima \tau &=& 0~,  \cr
 7\ima z - 6\ima \rho- 2\ima \tau &=& 0~, \cr
5 \ima z - 3\ima \rho - 2\ima \tau &=& 0  ~.
\eea
Higher values of $t$ work in a similar fashion. It is interesting to note that  the chambers for the paramodular groups with $t\neq1$ are the union of chambers for $t=1$, i.e. $SP(4,\ZZ)$, since the poles in  \eqref{eq:immaster} are a subset of those for that group. This means we will no longer have a tessellation of the Siegel upper half plane made of triangles, but rather of polygons with additional faces.  Figure \ref{chamberRpara} contrasts the chamber for $t=1$ versus $t=2$.

Note that linear poles in \eqref{eq:immaster} can be obtained from a subgroup of $\Gamma_t^+$, namely transformations of the form 
\be\label{eq:aa11}
\hat \gamma_t :=\twobytwo{{\bf A}}{0}{0}{{\bf D}}~, \quad {\bf A}{\bf D}^T=\mathds{1}_{2\times 2}~, \quad \det({\bf A})=\pm1~. 
\ee
Given the restrictions in \eqref{eq:defg2}, we can parametrize the matrix ${\bf A}$ as
\be\label{eq:aa1}
{\bf A} = \twobytwo{a_1}{t b_1}{c_1}{d_1}~ , \quad a_1d_1-t b_1c_1=\pm1~,
\ee
with $a_1,b_1,c_1,d_1\in\ZZ$. Note that ${\bf A} \in GL(2,\ZZ)$, and is allowed to have determinant -1.  
The linear pole \eqref{eq:immaster} corresponds in this notation to a $\hat \gamma_t$ transformation of $z=0$ where 
\be
a=a_1 c_1~,\quad b =a_1d_1+tb_1c_1~,\quad c= d_1b_1 ~.
\ee
  In addition, the extension in \eqref{eq:defgammat} allows for matrices ${\bf A}$  of the form
\be
\twobytwo{\sqrt{t}b_1}{\sqrt{t} a_1}{d_1/\sqrt{t}}{\sqrt{t}c_1}~,  \quad a_1d_1-t b_1c_1=\pm1~.
\ee
Relative to \eqref{eq:aa1}, these elements  swap the role of $\rho\to  t^{-1}\tau $ and $\tau\to t \rho$ as in \eqref{eq:flip}, and hence just gives redundant information in relation to \eqref{eq:aa1}. For this reason, we do not need to consider them in subsequent derivations.

\subsection{Crossing walls}

How can we now compute the degeneracy $c(n,m,l)$ of a charge vector $(n,m,l)$ in the chamber \textbf{R}? There is one simple special case where we can read off the result immediately: if  
\be\label{standardform}
n < -A \qquad \textrm{or} \qquad m < -C\ , 
\ee
then $c(n,m,l)=0$, as can be seen directly from expanding (\ref{explift2}). For our five examples in Sec. \ref{sec:five}, we have $A=1/t$ and $C=1$, and hence
\be\
c(n,m,l)=0~,\qquad {\rm for} \quad n < -{1\over t} \quad \textrm{or} \quad m < -1~.
\ee
We will call a vector $(n,m,l)$ which satisfies (\ref{standardform}) to be of \emph{standard form}. This immediately suggests a strategy to compute the degeneracy for an arbitrary charge vector as originally proposed in \cite{Sen:2011mh}: we can try to find an element $\hat\gamma\in\Gamma_t^+$ which transforms our charge vector to standard form. The price we pay for this is that $\hat{\gamma}$ permutes the chambers, so that we are no longer in \textbf{R}. We can then however deform the contour back to $\textbf{R}$, picking up contributions from all the poles we cross in the process. Since the charge vector is now in standard form, its contribution in $\textbf{R}$ vanishes, so that the degeneracy of the original charge vector is simply given by the sum of the residues of all the poles we crossed in going back to $\textbf{R}$. To be more explicit, let's define the matrix
\be
Q \equiv \twobytwo{n}{{l\over2}}{{l\over 2}}{m}~,
\ee
for which \eqref{cdef} takes the form
\be\label{eq:cdef1}
c(m,n,l)=\oint_{p=0} {dp\over 2\pi i p} \oint_{q=0} {dq\over 2\pi i q} \oint_{y=0} {dy\over 2\pi i y} \, \frac{e^{2\pi i{\rm Tr}(\Omega Q)}}{\Phi_k(\Omega)} \,.
\ee
Acting with $\hat \gamma$ in \eqref{eq:aa1} acts on our integration variable as
\be
\hat\gamma(\Omega)={\bf A}\Omega {\bf D}^{-1}\equiv \twobytwo{\tau_\gamma}{{z_\gamma}}{{z_\gamma}}{\rho_\gamma}\,,
\ee
which leads to a new charge vector in \eqref{eq:cdef1} of the form
\be\label{eq:gq1}
 {\bf D}Q {\bf A}^{-1} \equiv \twobytwo{n_\gamma}{{l_\gamma\over2}}{{l_\gamma\over 2}}{m_\gamma}~.
\ee
In components this reads  as
\bea
\tau_\gamma&=&a_1^2 \tau +2t a_1b_1 z + t^2b_1^2\rho ~, \notag \\
z_\gamma&=&a_1c_1 \tau + (a_1d_1+tb_1c_1) z + tb_1d_1 \rho~, \\
\rho_\gamma&=&c_1^2 \tau + 2 c_1d_1 z + d_1^2 \rho \notag ~,
\eea
and
\bea\label{eq:mnlg1}
m_\gamma&=&a_1^2 m + t^2b_1^2 n - ta_1 b_1 l~, \notag \\
n_\gamma&=& c_1^2 m + d_1^2n -c_1d_1 l ~,\\
l_\gamma&=& -2 a_1 c_1 m - 2t b_1 d_1 n +(a_1d_1+tb_1c_1) l \notag ~.
\eea
A transformation $\hat\gamma$ in \eqref{eq:gq1} gives a charge vector in standard form if
$m_\gamma < -1$ or $n_\gamma <-{1/t}$. Let us assume that this can be accomplished by some transformation $\hat{\gamma}_0$ that brings us to a chamber $\textbf{R'}\neq\textbf{R}$. It is left for us to trace our way back to the chamber \textbf{R}, picking up the contribution of the poles ${\bf p}_i$ that we cross along the way. The degeneracies will therefore take the form
\be
c(m,n,l)= \sum_{{\bf p}_i} \frac{1}{2\pi i}{\rm Res} \left({q^{-n}p^{-m}y^{-l}\over \Phi_k}, {\bf p}_i\right) \,.
\ee
Here ``{\rm Res}'' stands for the residue integral around $p_i$ and the two integrals for the remaining variables (for which we give a closed expression below).

To extract the residues at the poles, we will use the simple form of our paramodular forms near the pole $z=0$ as given in \rref{liftresidue}. Any linear pole ${\bf p}_i$ can be mapped to the pole $z=0$ by an appropriate transformation $\gamma_i \in \gamma_t$. From \eqref{eq:mnlg1}, and using \rref{liftresidue}, the residue at the pole $z_{\gamma_i}=0$ is given by
\be \label{polecontribution}
\frac{1}{2\pi i}{\rm Res}  \left({q^{-n}p^{-m}y^{-l}\over \Phi_k}, {\bf p}_i\right)= -l_{\gamma_i} d_t(n_{\gamma_i}) \tilde{d}_t(m_{\gamma_i}) \,,
\ee
where $d_t(n)$ and $\tilde d_t(n)$ are integers, whose relation to Dedekind-eta functions are   
\be
\eta(\tau)^{-24/t} = \sum_n d_t(n) q^n \,, \qquad  \eta(t \tau)^{-24/t} = \sum_n \tilde{d}_t(n) q^n~.
\ee
From this definition it is automatic that $d_t(n)$ and $\tilde d_t(n)$ vanish for   $n < -{1\over t}$ and/or $m < -1$.

A few comments are in order concerning our expression \rref{polecontribution}.
\begin{enumerate}
\item First, note that the factor of $l_{\gamma_i}$ in \eqref{polecontribution} comes from the fact that we picked $m_{1,1}=2$. This gives a  second order pole for which we can easily and explicitly cast the answer as in \eqref{polecontribution}. Higher values of $m_{1,1}$ are conceptually equivalent, and our discussion could be extended to those cases, with the caveat that the residue formula is more involved.    
\item Second, it is important to note that the contribution of a pole is conditional and only gives a non-vanishing result if
\be\label{eq:ccc3}
\text{Im} z_{\gamma_i} l_{\gamma_i} >0~.
\ee
This condition makes sure that we are crossing the pole from the right side. The element $\gamma_i$ must map us to the side of the pole away from \textbf{R}. 
\item Third, note that the contribution of a pole can vanish even if $\textbf{R'}\neq\textbf{R}$ if at ${\bf p}_i$ we have  $n_{\gamma_i}<-1/t$ or $m_{\gamma_i}<1$. This is not surprising because  there are arbitrarily many chambers in which the charge vector is in standard form: the procedure is not unique since we can map a given $(m,n,l)$ state to an arbitrarily negative $m_\gamma$ or $n_\gamma$ state via \eqref{eq:mnlg1}, taking us as deep as we want in the tessellation. The implication is that there are many paths back to \textbf{R}. However the Fourier coefficient $c(m,n,l)$ is of course unique, and this is compatible with the fact that a longer path will have several trivial contributions to the residue.
\item Finally,  one may wonder why this technique was not used to compute the Fourier coefficients of positive a discriminant states ($\Delta>0$) that are relevant for counting black hole microstates. The reason is simply because a   charge vector with positive discriminant can never be put in standard form via an element of $\Gamma_t^+$. This can be seen directly from the expression for $m_\gamma$ and $n_\gamma$ in \eqref{eq:mnlg1}. Our limitation by using  this elegant technique is to states with $\Delta<0$.
\end{enumerate}

To be very concrete about the implementation of the above ideas, let us study an example. We will consider $t=1$ and determine the Fourier coefficient $c(6,0,-3)$. From a direct expansion of $1\over\Phi_{10}$, we have
\be \label{cex}
c(6,0,-3)=1848528 \,.
\ee
Now consider the transformation $\gamma_0$ of the form \eqref{eq:aa11} where we select
\be
 {\bf A}_0 = \twobytwo{1}{-3}{0}{1} \,.
\ee
This gives
\be
n=0 ~\to~ n_{\gamma_0}=0 \,, \qquad m=6 ~\to~  m_{\gamma_0}=-3
\ee
The charge vector is now in standard form. Under this transformation, we have been sent from {\bf R} to the green region of Fig. \ref{chamberR}, and hence we need to cross the three poles to get back to {\bf R}:
\bea
{\bf p}_1&:& \quad z-3\rho=0~, \\ 
{\bf p}_2&:& \quad  z-2\rho=0~, \\ 
{\bf p}_3&:& \quad  z-\rho=0  \,.
\eea
 The contribution at the pole ${\bf p}_1$ vanishes:  the transformation $\gamma_1$ that maps the residue at ${\bf p}_1$ to $z=0$ vanishes since $\gamma_1=\gamma_0$ for which we would have $m_{\gamma_1}=-3$ in \eqref{polecontribution}. This is an illustration of our third point made above.  The contributions at the other two poles are non-trivial and read
\bea
\frac{1}{2\pi i}{\rm Res}  \left({q^0p^{-6}y^{3}\over \Phi_k}, {\bf p}_2\right) &=&1728~, \\
\frac{1}{2\pi i}{\rm Res}  \left({q^0p^{-6}y^{3}\over \Phi_k}, {\bf p}_3\right) &=&1846800~. 
\eea
Adding them up, we find
\be
c(6,0,-3)=1848528 \,,
\ee
in perfect agreement with \rref{cex}.

\subsection{The single pole regime}\label{sec:singlepole}

The prescription described in the previous subsection gives a constructive algorithm to compute the exact degeneracy for arbitrary charge vectors with negative discriminant. In this section, we will focus on a particular set of charge vectors, namely those whose Fourier coefficients are given by the contribution of a single pole. It will become clear in the following section that this restriction is relevant when trying to give a holographic interpretation to our results and describe supergravity spectra.
We will first work out the two examples $t=1,2$ in complete details to illustrate this, and then say a few words about the case $t>2$.

\subsubsection{Example: $t=1$}\label{sec:method1}

For $t=1$, the only pole bordering \textbf{R} from below is
\be
{\bf p}:\quad \ z-\rho=0 \,.
\ee
The contribution at that pole can be obtained from the element
\be
{\bf A}=\twobytwo{1}{-1}{0}{1} \,,
\ee
which gives
\be \label{cteq1}
c(m,n,l)=-(l+2n) d_1(n)d_1(m+n+l)\ .
\ee
We are looking for charge vectors for which $c(m,n,l)$ comes from (\ref{cteq1}) only. The most natural guess is of course to take a $(m,n,l)$ which is in standard form after the map ${\bf A}$. Note however that in that case, (\ref{cteq1}) immediately tells us that $c(m,n,l)$ vanishes, since by definition of the standard form either $d_1(n_\gamma)$ or $d_1(m_\gamma)$ vanishes. 

To get a non-vanishing answer, we therefore want a charge vector which is put in standard form by a transformation that maps ${\bf R}$ either into the red or yellow chamber in Fig. \ref{chamberR}, that is one chamber beyond the blue chamber. The elements $\hat \gamma_{1,2}$ that map to the red/yellow chambers respectively are
\be
{\bf A}_1=\twobytwo{1}{-2}{0}{1} \,, \qquad {\bf A}_2=\twobytwo{1}{-2}{-1}{1} \,.
\ee
Crossing into the blue chamber, the contribution of the pole then vanishes for the reasons given above. The total contribution that we pick up by deforming the contour back into ${\bf R}$ is thus simply (\ref{cteq1}), as desired.

The question is then of course for which regime of $n,m$ and $l$ this happens.
Using \eqref{eq:mnlg1} gives
\bea
&n_{\gamma_1}=n~, \quad
&m_{\gamma_1}=m+4n+2l~, \cr
&n_{\gamma_2}=m+n+l ~,\quad
&m_{\gamma_2}=m+4n+2l~.
\eea
For the charge vector to be in standard form in either of the chambers, we need one (or more) of these values to be less than minus one. Since we also want the Fourier coefficient \rref{cteq1} to be non-zero, the only possibility is
\be
m+4n+2l <-1 \,.
\ee
This defines the full one pole regime and the value of the coefficient is \rref{cteq1}.{\footnote{There is actually another pole that we could have crossed, the pole $z-\tau=0$. The reason we omitted this pole is because it is never relevant for charge vectors with $m>n$. Since we will ultimately be interested in the large $m$ limit, that other pole is never relevant. Note however that it would have been straight forward to add and for $t=1$, it would simply give equivalent formulas where $m$ and $n$ are exchanged.}

A formula for the degeneracy of negative discriminant in CHL compactifications was derived in \cite{Gomes:2015xcf} by exploring the black hole residue formula and the Rademacher expansion \cite{Dijkgraaf:2000fq}. 
We can check that equation (\ref{cteq1}) agrees with the result derived in \cite{Gomes:2015xcf}  after the appropriate change of variables. The method exploited here may provide a possible derivation of the contour choice made in \cite{Gomes:2015xcf}.

\subsubsection{Example: $t=2$}

Let us now look at the case $t=2$. This case is slightly more complicated since there are two poles bordering \textbf{R} from below. We have
\bea
{\bf p}_L&:& \quad z-2\rho=0 \\ 
{\bf p}_R&:& \quad  3z-2\rho-\tau=0 \,.
\eea
The elements that map ${\bf R}$ to the chambers delimiting these poles from above are
\be\label{t2Amap}
\hat{\gamma}_L: ~ {\bf A}_L=\twobytwo{1}{-2}{0}{1} \,, \qquad \hat{\gamma}_R: ~ {\bf A}_R=\twobytwo{1}{-2}{-1}{1}~,
\ee
which give
\bea \label{cteq2L}
c_L(m,n,l)&=&-(l+4n)d_2(n)\tilde{d}_2(m+4n+2l) ~,\\
c_R(m,n,l)&=&-(3l+2m+4n)d_2(l+m+n)\tilde{d}_2(m+4n+2l) \label{cteq2R} \,.
\eea
Let us denote by \textbf{C}$_{L,R}$ the two image chambers of ${\bf R}$ under the maps (\ref{t2Amap}), whose upper boundaries are of course given by ${\bf p}_{L,R}$. By the same reasoning as for $t=1$, to obtain non-vanishing coefficients, we need to consider transformations into chambers adjacent to \textbf{C}$_{L,R}$, of which there are three each.

For the pole ${\bf p}_L$, the transformations correspond to the elements
\be
{\bf A}_{L,1}=\twobytwo{1}{-4}{0}{1} \,, \qquad  {\bf A}_{L,2}=\twobytwo{1}{-4}{-1}{3} \,, \qquad {\bf A}_{L,3}=\twobytwo{1}{-2}{-1}{3} \,,
\ee
which give
\bea
&n_{\gamma_{L,1}}=n ~,\quad 
&m_{\gamma_{L,1}}=m+16n+4l~, \cr
&n_{\gamma_{L,2}}=m+9n+3l ~,\quad
&m_{\gamma_{L,2}}=m+16n+4l~,\cr
&n_{\gamma_{L,3}}=m+9n+3l ~,\quad
&m_{\gamma_{L,3}}=m+4n+2l~.
\eea
For the pole ${\bf p}_R$, the transformations correspond to the elements
\be
{\bf A}_{R,1}=\twobytwo{1}{-2}{-2}{3} \,, \qquad  {\bf A}_{R,2}=\twobytwo{3}{-4}{-2}{3} \,, \qquad {\bf A}_{R,3}=\twobytwo{3}{-4}{-1}{1} \,,
\ee
which give
\bea
&n_{\gamma_{R,1}}=4m+9n+6l ~,\quad 
&m_{\gamma_{R,1}}=m+4n+2l ~, \cr
&n_{\gamma_{R,2}}=4m+9n+6l ~,\quad 
&m_{\gamma_{R,2}}=9m+16n+12l ~, \cr
&n_{\gamma_{R,3}}=m+n+l ~,\quad 
&m_{\gamma_{R,3}}=9m+16n+12l~.
\eea
Putting everything together, the degeneracies \rref{cteq2L} and \rref{cteq2R} are valid  when 
\bea
\text{Regime L}&:& \ \ m+16n+4l < -1/2  \ \cup \  m+9n+3l <-1~,\\
\text{Regime R}&:& \ \ 4m+9n+6l < -1/2  \ \cup \ 9m+16n+12l<-1~,
\eea
respectively. These regimes exclude as well ranges of $(m,n,l)$ for which the Fourier coefficients are trivial.

\subsubsection{$t>2$}
Let us now discuss the general case.
In principle this is straightforward, but rather cumbersome to write down explicitly. The reason is that the chamber \textbf{R} is bounded below by $2^{t-1}$ poles as can be seen for example in Fig. \ref{teq2and3}. This gives $2^{t-1}$ chambers below {\bf R} which are each bounded below by $2^{t-1}+1$ poles. This gives in principle $2^{t-1}(2^{t-1}+1)$ different paths from a chamber where the charge vector is in standard form back to \textbf{R}, each of  which gives a different regime for the charge vector. We will therefore concentrate on just two such paths.

The leftmost pole closest to ${\rm Im}z=0$ is given by
\be
{\bf p}_L:\quad \ z-t\rho=0 ~.
\ee
We will now only consider the transformation that maps to the leftmost chamber adjacent to this chamber (This is of course not the only path, as there are another $2^{t-1}$ other poles that we could cross. This means the regime that we will write down shortly is not the largest possible regime where the pole ${\bf p}_L$ gives the only contribution).
The relevant element in \eqref{eq:aa11} is then
\be
{\bf A}_0 = \twobytwo{1}{-2t}{0}{1} \,.
\ee
The charge vector will be put in standard form provided
\be \label{regimepL}
m+4t^2n+2t l < - 1 \,,
\ee
and only the pole ${\bf p}_L$ will contribute. The answer yields
\be \label{cLonepole}
c_L(m,n,l)=-(l+2 t n) d_t(n)\tilde{d}_t(m+t^2n+t l) \,.
\ee
This formula is exact provided \rref{regimepL} is satisfied.

The rightmost pole is given by
\be
{\bf p}_R: \quad (2t-1)-t\rho -(t-1)\tau=0 \,.
\ee
Note that for $t=1$ this gives the same pole, which is consistent since \textbf{R} is bounded below only by one pole in that case. To put the vector in standard form, we use the element
\be
{\bf{A}}_0 = \twobytwo{2t-1}{-2t}{-1}{1} \, ,
\ee
which maps to the rightmost adjacent chamber.
The charge vector will be put in standard form if
\be \label{regimepR}
(4t^2-4t-1)m+4n+2t(2t-1)l < - 1 \,,
\ee
and only the pole ${\bf p}_R$ will contribute. The answer reads
\be
c_R(m,n,l)=-(2(t-1)m+(2t-1)l+2tn) d_t(l+m+n)\tilde{d}_t((1-2t+t^2)m+t^2n+t(t-1)l) \,.
\ee
This expression is again exact provided \rref{regimepR} is satisfied. It is important to note that both regimes \rref{regimepL} and  \rref{regimepR} are necessary but not sufficient: there are additional negative discriminant states   whose Fourier coefficient is equal to the non-trivial residue at one pole. These additional states are those brought to standard form by considering the other $2^{t-1}$ paths that we ignored here as we cross the first pole  and then there are also $2^{t-1}-2$ middle poles to reach back to {\bf R}. 

We will now apply the method we developed to extract the Fourier coefficients of symmetric product orbifold theories. We will see that the contribution of a single pole contributing has a nice interpretation from a holographic perspective.

\section{Symmetric orbifolds \& Siegel paramodular forms}\label{sec:sfl}

In this section we discuss our main application of Siegel paramodular forms:  quantifying  the growth of BPS operators for supersymmetric CFTs coming from symmetric orbifolds. Given a seed theory ${\cal C}$, we construct a symmetric product orbifold by tensoring $r$ copies of the seed and then orbifolding by the symmetric group $S_r$, giving 
\be
\mathcal{C}_r \equiv \frac{\mathcal{C}^{\otimes r}}{S_r} \,.
\ee
Assuming that the seed theory has an elliptic genus of index $t$ whose coefficients are given by $c(n,l)$,
the generating function for the elliptic genus of the $r$-th orbifolded theory is \cite{Dijkgraaf1997,Dijkgraaf1998,Bantay2003} 
\be \label{ZSymN}
\mathcal{Z}(\rho,\tau,z)=\sum_{m\in t \mathbb{N}}p^m \chi_m(\tau,z)=\prod_{\substack{n,l,r\in\ZZ
		\\r>0}} \frac{1}{(1-q^n y^l p^{tr})^{c(nr,l)}} \,.
\ee
Here $ \chi_{tr}(\tau,z)$ is  the elliptic genera that captures BPS states of $\mathcal{C}_r$ with Fourier coefficients defined by
\be\label{eq:cm1}
 \chi_m(\tau,z)=\sum_{n\geq 0,l\in \ZZ} c_{\rm CFT}(m,n,l) q^n y^l~.
\ee

The generating function $\mathcal{Z}$ is closely related to a Siegel paramodular form. Focusing on our five examples in Sec. \ref{sec:five},  from \eqref{explift3} we have
\be \label{relationSMFSymN}
\mathcal{Z}(\rho,\tau,z)=\frac{p\,\phi_{k,1}(\tau,z)}{\Phi_k(\rho,\tau,z)} \,,
\ee
where the Hodge factor is
\be\label{Hodgephikt}
\phi_{k,1}(\tau,z)=q^{1/t} y \prod_{\substack{(n,l)>0} }(1-q^n y^l)^{c(0,l)} \, ,
\ee
and $\Phi_k$ is a Siegel paramodular form given by an exponential lift of the form \eqref{explift2}; in \eqref{relationSMFSymN} we used \eqref{eq:abc1}--\eqref{eq:m112}.
It is therefore clear how to extract the Fourier coefficients of the symmetric product once those of the Siegel paramodular form are known. In the following we will discuss the interpretation of the negative discriminant states in ${1/\Phi_k}$ we quantified in Sec. \ref{sec:method} in relation to states in ${\cal Z}$.

Despite their close relation, it is worth highlighting some differences in the states contained in ${\cal Z}$ relative to ${1/\Phi_k}$.  In the expansion of ${\cal Z}$, the coefficient of $p^m$ is a weak Jacobi form of index $m$, which has polar states, i.e. states with $\Delta=4mn-l^2 <0$. For fixed index $m$, the polar states are bounded from below by  $\Delta \geq - m^2$ as expected. In contrast, the expansion of ${1/\Phi_k}$ in powers of $p$ are not weak Jacobi forms, and this leads to having in its expansion negative discriminant states that are not bounded by $m$.\footnote{This is a simple consequence of the expansion of ${1/\Phi_k}$ around the pole $y=1$: this allows for arbitrarily large positive powers of $l$.} The  Hodge factor (\ref{Hodgephikt}) is what reconciles the expansion on both sides  of \eqref{relationSMFSymN}, and the discriminants of states on both sides.  Moreover, $\phi_{k,1}$  only contains positive discriminant states: this means that to understand the polar states in (\ref{relationSMFSymN}) it is enough to know the negative discriminant states of $\Phi_k$.\footnote{The tensor product of two states of positive discriminant always results in a state with positive discriminant. This follows from the `Lorentzian' triangle inequality
	$
	||x+y||>||x||+||y||
	$
	where $||\ldots||$ is the $SO(1,2,\mathbb{R})$ invariant norm and $x,y$ denote states $(m,n,l)$ with $m>0$. Hence a $\Delta>0$ in ${1/\Phi_k}$  is  a non-polar state in ${\cal Z}$.} 

In the remainder of this section we will study the degeneracies of polar states in ${\cal Z}$. Our emphasis will be on identifying those polar states which we can interpret holographically as perturbative states of a putative theory of gravity on AdS$_3$. This will require a definition of vacuum state (and performing  a suitable spectral flow to identify it), in addition to a notion of lightness in the CFT which we discuss in the following subsection.  

\subsection{Light operators}
In the following we will discuss the physical interpretation of $\chi_m$ in \eqref{ZSymN}, and its operator content, with particular emphasis on `light' operators which we define below. 
Our five examples of Siegel paramodular forms involve as a seed a weak Jacobi form of weight zero. The most natural interpretation of a weak Jacobi form is as the elliptic genus of a theory with $N=(2,2)$ supersymmetry,\footnote{It is important to highlight that our interpretation of $\chi_m$ is not limited to $N=(2,2)$ theories. Our requirements are that the theory has some amount of supersymmetry, an $R$-symmetry and that a weak Jacobi form is the relevant mathematical object that counts BPS states. We use the $N=(2,2)$ SCFT jargon for concreteness. } and hence we would identify 
\be
\chi_m(\tau,z)=\tr_{RR}\le((-1)^{F} q^{L_0-\frac{c}{24}} y^{ J_0 } \bar{q}^{\bar{L}_0-\frac{\bar{c}}{24}} \ri)~.
\ee
As denoted by the subscript, this index is defined in the Ramond sector where fermions have periodic boundary conditions.   $L_0$ and $\bar L_0$ are the zero modes of the  left and right Virasoro generators; there are also two $U(1)$ R-charge operators with zero modes  $J_0$ and $\bar J_0$.\footnote{The $R$-symmetry of the SCFT$_2$ can be larger than $U(1)$; it could be for instance $SU(2)$ or larger. For the purpose of our argument we need to just focus on the $U(1)$ subgroup of the appropriate group.}  With the fermion number given by $F=J_0-\bar J_0$, the insertion of $(-1)^{F}$ turns the resulting object into a holomorphic function. Following the notation in \eqref{eq:cm1}, $(n,l)$ are  the eigenvalues of $L_0-c/24$ and $J_0$ respectively. 

Since the elliptic genus is defined in the Ramond sector, it does not count perturbative low-energy states in AdS: in particular we do not expect the vacuum state to be in this sector. To address this issue let us first discuss some features of the CFT associated to $\chi_m$, and in particular the vacuum state. Consider the left moving sector of a  $N=(2,2)$ SCFT$_2$. We will focus on the Virasoro algebra of that sector and the $U(1)$ Kac-Moody algebra due to the $R$-symmetry.  The relevant commutators are
\bea
\label{eq:canonicalgebra}
[ L_n, L_{n'}] &=&(n-n') L_{n+n'}+\frac{c}{12}n(n^2-1)\delta_{n,-n'}~, \cr
[ L_n, J_{n'}] &=&-n'  J_{n'+n}~, \cr
[ J_n, J_{n'}] &=&2m n\,\delta_{n,-n'}~.
\eea
In a superconformal theory, the level (index) $m$ is related to the central charge $c$ via $c=6m$; 
but for now we will keep them unrelated. The additional generators and properties of a superconformal algebra can be found in, for example, \cite{Blumenhagen:2009zz}. For our purposes, an important feature is the invariance of the algebra under a continuous family of deformations, known as a spectral flow automorphism \cite{Schwimmer:1986mf}:
\bea
\label{eq:spectralflow}
 L_n &\to & L_n^{(\rm sf)}= L_n+\eta\,  J_n+{\eta^2} \,m\, \delta_{n,0}~,\cr
 J_n &\to & J_n^{(\rm sf)}=J_n +2\eta \,m\, \delta_{n,0}~.
\eea
Under this deformation the elliptic genus transforms as
\be
 \chi_m(\tau,z)\quad\to \quad \chi_m^{\rm sf}(\tau,z)=q^{\eta^2m} y^{2\eta m} \chi_m(\tau,z+\eta\tau) \,.
\ee
Here $\eta$ is a continuous parameter. In particular, for $\eta\in \ZZ +1/2$ the deformation interpolates between the R sector (periodic fermions) and the NS sector (anti-periodic). The case $\eta\in \ZZ$, corresponds to  the  translation in \eqref{eq:jf2}.  Note that the discriminant, $\Delta=4nm-l^2$, is invariant under \eqref{eq:spectralflow}.

The vacuum state is defined as a highest weight state whose zero modes are
\be\label{eq:vac1}
L_0 | 0\rangle = 0  ~, \quad J_0 | 0\rangle =  0~,
\ee
and it is annihilated by $L_{-1}$ and $(L_n,J_n)$ with $n>0$. Equivalently, the vacuum is a state invariant under the $sl(2)\times u(1)$ subgroup in \eqref{eq:canonicalgebra}, and has the lowest value of the discriminant: $\Delta_{\rm vac}=-{cm\over 6}$. It is important to note that \eqref{eq:vac1} is not spectral flow invariant.  This is the reason why  the weak Jacobi form $\chi_m(\tau,z)$ does not count the vacuum: there are no negative powers of $q$ in \eqref{eq:cm1}.

Our task now is to find a suitable spectral transformation such that $\chi_m^{\rm sf}$ contains a state of the form \eqref{eq:vac1}. However, how we flow to the sector containing the vacuum depends on how we relate the index $m$ of the elliptic genus to the central charge. One natural interpretation is to simply take the seed weak Jacobi form to be the elliptic genus of a SCFT with central charge
\be
c_{\rm seed}=6t \,,
\ee
as would be the case for the elliptic genus of a Calabi-Yau sigma model, i.e. the interpretation given in  \cite{Gritsenko:1999fk} \footnote{To be more precise the central charge of a Calabi-Yau (CY$_d$) sigma model is given by $c=3d$ with $d$ the compex dimension. In addition, we also have that the index of the elliptic genus is given by $d/2$. When $d$ is even we have $t=d/2$. }. The vacuum is then in the NS sector, that is in the sector that is obtained by spectral flow by a half unit $\eta =\frac{1}{2}$. For this choice of central charge the seed theory has
\be
\Delta_{\rm vac}=-t^2~,
\ee
which is the most polar term allowed by the index of the seed. However, for our five examples in Sec. \ref{sec:five} the minimal polarity of the seed is $\Delta_{\rm min}=-1$. 
 If $t=1$, that is the $K3$ sigma model SCFT, there is no tension and as we will see in the following section the analysis is rather straight forward.  However, for $t>1$ there are various issues. 
 The mismatch, $\Delta_{\rm vac}\neq\Delta_{\rm min}$,  means that the vacuum does not appear in $\chi_m^{\rm sf}$. As we review in Appendix \ref{calabiyau}, this can indeed happen for certain types of Calabi-Yau sigma models.  In addition there are  no light states --as defined below in (\ref{lightness1})-- in the NS sector, since the vacuum and its neighbors do not appear.  We will show this in Sec.~\ref{ss:halfint}.

A different approach is to set 
\be
\Delta_{\rm vac}= \Delta_{\rm min} =-1
\ee
which sets 
\be
c_{\rm seed}={6\over t}~.
\ee
The idea is to declare that the most polar term in the elliptic genus should be interpreted as the vacuum.  For this term to produce an uncharged state, as in \eqref{eq:vac1}, we see from \eqref{eq:spectralflow} that we need to flow by $\eta = \frac{1}{2t}$ units of spectral flow. We will call this choice the fractional spectral flow, and  we will study the spectrum of this sector in Sec.~\ref{ss:fracflow}.

We can summarize our two choices as:

\noindent\begin{minipage}{\textwidth}
\vspace{1em}
\begin{center}
\begin{tabular}{|c|c|c|c|}
\hline
& Vacuum maps to & central charge & $\eta$ \cr
\hline\hline
Half-Integer SF& Most polar term allowed by index & $c_{\rm seed}=6t$ & $1/2$ \cr
\hline
Fractional SF& Most polar term that is non-zero & $c_{\rm seed}=6/t$ & $1/(2t)$ \cr
\hline
\end{tabular}
\label{table:2choices}
\end{center}
\vspace{.2em}
\end{minipage}

Having these two choices, we can now properly define a light operator. 
A light state, or equivalently a perturbative state, is a  state whose weight is sufficiently small relative to the vacuum state. If we take $h$ to be the weight of the state above the vacuum, our definition of  light states is to require
\be \label{lightness1}
\textbf{Lightness :} \qquad \frac{h}{c} \to 0 \,, \quad c\to \infty \,.
\ee
That is, our definition of lightness comes intrinsically with a large central charge limit; this naturally takes us to the corner of holography where the states in AdS are perturbative fields in the supergravity description.  For our five examples, we have $c=mc_{\rm seed}$, and hence \eqref{lightness1} is a large $m$ limit. 

Since $h$ is the useful variable to quantify energy, we will write
\be
q^{c/24}\chi_m^{\rm sf}(\tau,z) =: \sum_{h,l_{\rm sf}}c_{\rm sf}(m,h,l_{\rm sf})q^h y^{l_{\rm sf}}~,
\ee
where
\be\label{SF}
h= n + \eta l ~, \qquad l_{\rm sf} = l + 2m\eta\ .
\ee
and the shift by the vacuum energy ($-c/24$) is taken into account.

 In the remainder of this section we will evaluate generating functions for $c_{\rm sf}(m,h,l_{\rm sf})$ for polar states that satisfy  the single pole regime introduced in Sec. \ref{sec:singlepole}. A priori there is no evident reason why one should focus on those states. However, as our computations will reveal, these states are exactly those relevant to discuss the light regime in  \eqref{lightness1}. Of course one could build the generating function for all polar states in $\chi_m^{\rm sf}(\tau,z) $, but our primary task in this portion is to establish that our five examples have a very sparse spectrum as defined by \eqref{eq:cS1}.

\subsection{Example: $t=1$}
The simplest case is $t=1$, where $\chi_m$ is interpreted as the elliptic genus of $K3$. The seed of this theory is given in \eqref{eq:p01} and the paramodular form in \eqref{eq:t1k10}. The light states  are in the NS sector, which we reach by a $1/2$ unit spectral flow. The parameters as defined in \eqref{SF} read in this case
\be\label{eq:nh1}
n = h - \frac{1}{2}l_{\rm NS} \qquad l = l_{\rm NS} - m \ .
\ee

To start we will implement the spectral form on the Siegel form $1/\Phi_{10}$, and later on introduce the hodge factor to obtain ${\cal Z}$ via \eqref{relationSMFSymN}.  We define the coefficients in the NS sector as 
\be
c_{\rm NS}(m,h,l_{\rm NS}):= c(m,n,l)=c\left(m,h-\frac{l_{\rm NS}}{2},l_{\rm NS}-m\right)\ .
\ee
We can now define $1/\Phi_{10}$ in the NS sector as
\be
\frac{1}{\Phi^{\rm NS}_{10}}= \sum_{m,h,l_{\rm NS}} c_{\rm NS}(m,h,l_{\rm NS}) p^m q^h y^{l_{\rm NS}}\ .
\ee

We could like to characterise $c_{\rm NS}(m,h,l_{\rm NS})$ based on our findings in Sec. \ref{sec:method}. In particular, we will build a generating function in the NS sector for all polar states that lie in the single pole regime; these are the states quantified in Sec. \ref{sec:method1}. The condition of the single pole regime for $t=1$ is given by  \rref{cteq1}, which in NS variables reads
\be \label{singleresidue}
4h-m \leq -2 \,.
\ee
Provided this conditions is satisfied we have \eqref{cteq1}, which in NS variables maps to
\be\label{CNSs}
c_{\rm NS}(m,h,l_{\rm NS})= (m-2h)d(h-\frac{l_{\rm NS}}{2})d(h+\frac{l_{\rm NS}}{2})\ .
\ee
Let us define $c_{\rm NS}^{s}$ to be equal to (\ref{CNSs}), regardless of condition (\ref{singleresidue}), and define a generating function for them:
\bea\label{ZsNS}
Z^s_{\rm NS}(\tau,\rho,z) &\equiv&\sum_{h\geq-1} \sum_{|l_{\rm NS}|\leq 2h+2} \sum_{m\geq 2h+1}   (m-2h) d(h+\frac{l_{\rm NS}}{2})d(h-\frac{l_{\rm NS}}{2}) q^h y^{l_{\rm NS}}p^m \,.
\eea
We have chosen the summation range so that \eqref{CNSs} is compatible with  \eqref{eq:ccc3} and the entries of $d(n)$ are non-zero in \eqref{CNSs}, i.e.
\bea\label{eq:range1}
m-2h &\geq& 1 \notag \\
h-\frac{l_{\rm NS}}{2} & \geq & -1 \\
h+\frac{l_{\rm NS}}{2} & \geq & -1 \,.\nonumber
\eea
We are guaranteed that $Z^s_{\rm NS}$ agrees with the actual NS generating function $1/{\Phi^{\rm NS}_{10}}$ for terms which satisfy (\ref{singleresidue}).

$Z^s_{\rm NS}$ is  useful because it can be written in very simple form:
we can perform the sums in (\ref{ZsNS}) to find
\bea\label{eq:zz1}
Z^s_{\rm NS}(\tau,\rho,z) &=&\sum_{h\geq-1} \sum_{|l_{\rm NS}|\leq 2h+2}d(h+\frac{l_{\rm NS}}{2})d(h-\frac{l_{\rm NS}}{2}) \left(qp^2\right)^h y^{l_{\rm NS}} \sum_{m'\geq 1}   m' p^{m'} \notag \\
&=&\frac{p}{(1-p)^2} \sum_{h\geq-1} \sum_{|l_{\rm NS}|\leq 2h+2}d(h+\frac{l_{\rm NS}}{2})d(h-\frac{l_{\rm NS}}{2}) \left(qp^2\right)^h y^{l_{\rm NS}} \notag \\
&=&\frac{p}{(1-p)^2} \sum_{r,s\geq-1} d(r)d(s) q^{(r+s)/2}p^{r+s}y^{r-s} \notag \\
&=& \frac{p}{(1-p)^2} \frac{1}{\eta(\tau/2+\rho+z)^{24}}\frac{1}{\eta(\tau/2+\rho-z)^{24}}~.
\eea
We are actually interested in the growth of coefficients in the symmetric orbifold $\mathcal{Z}$, which is the object we would want to match to the supergravity spectrum. It is easily extracted from $Z_{\rm NS}$ through \rref{relationSMFSymN}: we simply multiply \eqref{eq:zz1} by the spectrally flowed version of $p \phi_{10,1}$, which is the Hodge factor for the Igusa cusp form. We have
\be\label{phi101}
p \phi_{10,1}= p qy(1-y^{-1})^2 \prod_{n\geq1} (1-q^n)^{20}(1-q^n y)^2(1-q^n y^{-1})^2 \,.
\ee
We then do a half unit spectral flow transformation
\be
y \to y q^{1/2} \,, \qquad p\to p q^{1/2} y~,
\ee
and obtain
\be\label{HodgeK3}
p \phi^{\rm NS}_{10,1}= p q \prod_{n\geq1} (1-q^n)^{20}(1-q^{n-1/2}y)^2(1-q^{n-1/2}y^{-1})^2\ .
\ee
The single pole generating function of the symmetric orbifold is thus
\begin{align}\label{ZSymNS}
{\cal Z}^s_{\rm NS}&= Z^s_{\rm NS}\cdot  p \phi^{\rm NS}_{10,1}\cr&=\frac{1}{(1-p)^2} \prod_{n\geq1} \frac{(1-q^n)^{20}(1-q^{n-1/2}y)^2(1-q^{n-1/2}y^{-1})^2}{(1-q^{n/2}p^ny^n)^{24}(1-q^{n/2}p^ny^{-n})^{24}} \,.
\end{align}
To determine the regime of validity of (\ref{ZSymNS}), note that $Z^s_{\rm NS}$ is valid as long as (\ref{singleresidue}) is satisfied. Multiplying by (\ref{HodgeK3}) then changes this regime only slightly: any term that satisfies
\be\label{lightNSK3}
h \leq (m+1)/4 \,,
\ee 
necessarily comes from a term for which $Z^s_{NS}$ is valid. The difference to (\ref{singleresidue}) comes from the prefactor $pq$ in (\ref{HodgeK3}). All other factors in (\ref{HodgeK3}) only contain positive powers of $q$, which means that states in $Z^s_{NS}$ which violate (\ref{singleresidue}) never contribute to states in (\ref{ZSymNS}) which satisfy (\ref{lightNSK3}). It is also clear that all light states, as defined in \eqref{lightness1}, are contained within \eqref{lightNSK3}. It is rather interesting that the single pole regime is the natural regime to describe perturbative states.

 As demanded by \eqref{lightness1}, we are interested in studying the large central charge limit. We can extract the $m\to\infty$ limit of this product by essentially stripping off the $p=1$ pole \cite{deBoer:1998us, Keller:2011xi} and setting $p=1$. We find
\be\label{eq:zinf1}
{\cal Z}^\infty_{\rm NS}(\tau, z)= \prod_{n\geq1} \frac{(1-q^n)^{20}(1-q^{n-1/2}y)^2(1-q^{n-1/2}y^{-1})^2}{(1-q^{n/2}y^n)^{24}(1-q^{n/2}y^{-n})^{24}} \, .
\ee
Note that this result is exact now, since in this limit all states satisfy the bound \eqref{lightNSK3}, and hence ${\cal Z}^\infty_{\rm NS}$ counts all light BPS states in this case. To compare this to \cite{Benjamin:2015vkc}, we can further specialize to $y=1$; from \eqref{eq:zinf1} we have
\be
 \prod_{n\geq 1} \frac{1}{(1-q^n)^{28}(1-q^{n-1/2})^{44}}\ ,
\ee
which agrees with (39) in \cite{Benjamin:2015vkc}.\footnote{A derivation of \eqref{eq:zinf1} can also be found in \cite{Aharony:1999ti}, albeit their expression has a typo.}

\subsection{Half-integer spectral flow}\label{ss:halfint}

We now turn to analyzing the spectrum of $\chi_m^{\rm sf}$ for the half-integer spectral flow for $t>1$: this is the NS spectrum of the theory. Under ideal conditions, the half-integer spectral flow takes the term $p^mq^0y^m$ in the Ramond sector, i.e. $\chi_m$,  to the vacuum term in the NS sector. Unfortunately, for $t>1$ the Siegel paramodular forms start with $p^mq^0y^{m/t}$, and hence the most polar term does not appear in the counting formula. Despite this unappealing feature, it is worth describing the spectrum of $\chi_m^{\rm sf}$.

The steps we will take will mimic those for $t=1$: we will flow the Siegel paramodular form to the NS sector, and build a generating function for the states described in the single pole regime. The spectral flow and the shift of $h$ is identical as in \eqref{eq:nh1}, giving 
\be
h=n+\frac{l}{2}+\frac{m}{2} \,, \qquad l_{\rm NS}=l+m \,.
\ee
Let us start out by considering states which are in the regime (\ref{regimepL}), coming from the pole $p_L$ as described in Sec. \ref{sec:singlepole}. The single pole condition in the NS sector reads 
\be \label{regimegeneraltpL}  
(1-2t)m + 4t^2h+2t(1-t)l_{\rm NS} \leq-2  \,.
\ee
For such states their Fourier coefficients are
\be\label{cTt2}
c_{\rm NS}^{p_L}(m,h,l_{\rm NS})= (m-2th-(1-t)l_{\rm NS})d_t(h-\frac{l_{\rm NS}}{2})d_t(m(\frac{1}{t}-1) +ht+\lns(1-\frac{t}{2}))\ .
\ee
The formula \rref{cTt2} will be non-zero if the following three conditions are satisfied
\bea
 m-2th-(1-t)l_{\rm NS} &\geq& 0~, \notag \\
h-\frac{l_{\rm NS}}{2} &\geq& -1/t~, \\
m(\frac{1}{t}-1) +ht+\lns(1-\frac{t}{2}) & \geq & -1/t~. \notag
\eea
 Let us now write the generating function for the $c_{NS}^{p_L}(n,m,l)$, we have
\bea
Z_{\rm NS}^{p_L}(\tau,\rho,z)&=&\sum_{h,\lns,m} c^{p_L}_{\rm NS}(n,m,l) q^h y^{\lns} p^m ~,
\eea
where the sum is constrained by the three conditions above. To make the conditions more manifest, we will make a change of variables and set
\be
m'=-( \lns(1-t)-m+2t h)  \,, \quad r= h-\lns/2 \,, \quad s=m(1/t-1) +ht+\lns(1-t/2) \,,
\ee
which then gives
\be
Z_{\rm NS}^{p_L}=\sum_{m'\geq0, r,s\geq -1/t} m' d_A(r)d_A(s) q^{ts/2+m' (t-1)/2 +r(1+t(t-2)/2)} y^{ts+m'(t-1) +rt(t-2)}p^{ts+t m' +t^2 r} \,.
\ee
We now perform the sum over $m'$ and obtain
\be\label{eq:znst}
Z_{\rm NS}^{p_L}= \frac{p^t q^{(t-1)/2} y^{t-1}}{(1-p^t q^{(t-1)/2} y^{t-1})^2} \frac{ 1}{\eta(t\tau/2+t\rho+tz)^{24/t}} \frac{1}{\eta(\tau(1+t/2(t-2))+t^2\rho+t(t-2)z)^{24/t}} \,.
\ee
As noted around \eqref{eq:range1},  $Z_{\rm NS}^{p_L}$ will  match with the Fourier coefficients of $1/\Phi_k^{\rm NS}$ provided \eqref{regimegeneraltpL} is satisfied. However, \eqref{eq:znst} is a convenient intermediate object for studying the large $m$ limit.

To obtain the generating function of the symmetric product, we again need to add in the weighted part. Before spectral flow, we have
\be \label{phitgen}
p \,\phi_{k,1}= p q^{1/t}y(1-y^{-1})^2 \prod_{n\geq1} (1-q^n)^{24/t - 4}(1-q^n y)^2(1-q^n y^{-1})^2 \,.
\ee
The spectral flow transformation reads
\be
y \to y q^{1/2} \,, \qquad p\to p q^{1/2} y \,,
\ee
which gives
\be
p \,\phi^{\rm NS}_{k,1}= p q^{1/t} \prod_{n\geq1} (1-q^n)^{24/t - 4}(1-q^{n-1/2} y)^2(1-q^{n-1/2} y^{-1})^2 \,.
\ee
We can now obtain the final expression, which reads
\be\label{tstandardSF}
{\cal Z}_{\rm NS}^{p_L}=\frac{1}{(1-p^t q^{(t-1)/2} y^{t-1})^2} \prod_{n\geq1} \frac{(1-q^n)^{24/t - 4}(1-q^{n-1/2} y)^2(1-q^{n-1/2} y^{-1})^2}{(1-q^{tn/2}p^{tn}y^{tn})^{24/t} (1-q^{(1+t(t-2)/2)}p^{t^2}y^{t(t-2)})^{24/t}}~.
\ee
Note that this expression only gives the correct multiplicities for states in the regime
\be \label{regimegeneraltSymN}
(1-2t)m +4t^2h+2t(1-t) l_{\rm NS} \leq 2t-1 \,,
\ee
which came from \rref{regimegeneraltpL} after taking into account the shift from the weighted part. As for $t=1$, the difference between (\ref{regimegeneraltpL}) and (\ref{regimegeneraltSymN}) comes from the prefactor $pq^{1/t}$ in (\ref{phitgen}). Similarly one can check that the other factors in (\ref{phitgen}) never turn a state which violates (\ref{regimegeneraltpL}) into a state that satisfies (\ref{regimegeneraltSymN}).

It is important to note that the formula above mostly involves states with $l_{\rm NS}>0$, which tend to satisfy (\ref{regimegeneraltSymN}) more easily. 
We know in particular that the full answer for the symmetric product orbifold needs be invariant under $l_{\rm NS} \to - l_{\rm NS}$. The negative $l_{\rm NS}$ terms come in fact from the pole $p_R$. 
For the pole $p_R$, we gave the single residue in \rref{regimepR} which in NS variables reads
\be \label{regimegeneraltpR}
(1-2t)m+4t^2h-2t(1-t)l_{\rm NS} \leq -2~.
\ee
For such states, the degeneracies in the NS sector are 
\be
c_{\rm NS}^{p_R}(m,h,l_{\rm NS})= (m-2th-(t-1)l_{\rm NS})d_t(h+\frac{\lns}{2})d_t(m(\frac{1}{t}-1) +ht-\lns(1-\frac{t}{2}))\ .
\ee
Note that both the regime and the expression for the Fourier coefficients of $p_R$ are equal to that of the pole $p_L$ but with $\lns\to-\lns$. The non-vanishing states that are in those two regions are plotted in figure~\ref{f:lightstates} for $t=2$. 
\begin{figure}[htbp]
	\begin{center}
		\includegraphics[width=0.5\textwidth]{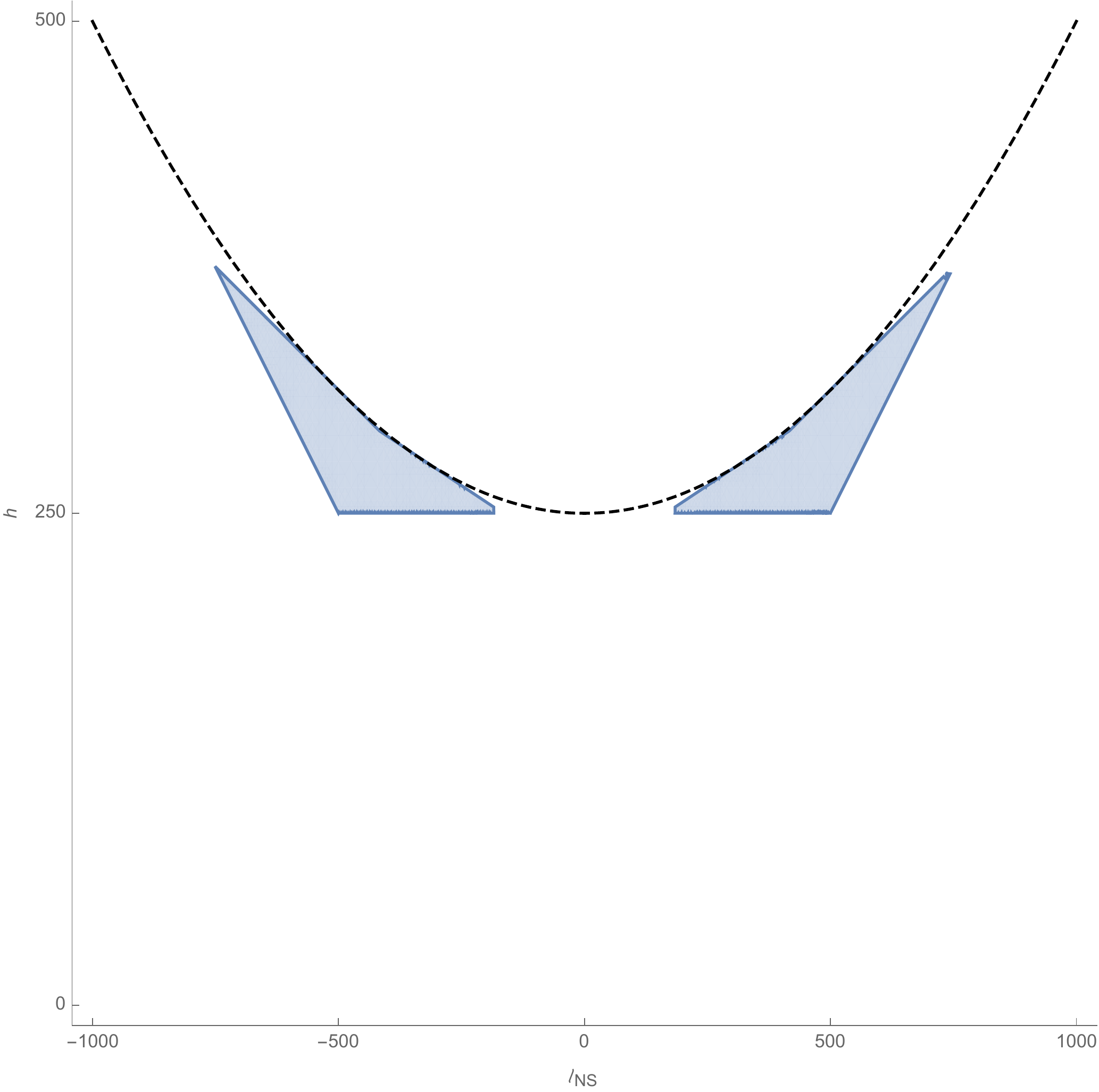}
		\caption{Non-vanishing states in region (\ref{regimegeneraltpL}) and (\ref{regimegeneraltpR}) for $t=2$ and $m=1000$. The dashed line is the boundary between polar and non-polar states.}
		\label{f:lightstates}
	\end{center}
\end{figure}
We may again compute the generating function for the symmetric product and we find
\be\label{tstandardSFpR}
{\cal Z}_{\rm NS}^{p_R}=\frac{1}{(1-p^t q^{(t-1)/2} y^{-(t-1)})^2} \prod_{n\geq1} \frac{(1-q^n)^{24/t - 4}(1-q^{n-1/2} y)^2(1-q^{n-1/2} y^{-1})^2}{(1-q^{tn/2}p^{tn}y^{-tn})^{24/t} (1-q^{(1+t(t-2)/2)}p^{t^2}y^{-t(t-2)})^{24/t}} \,,
\ee
which is only exact in the regime
\be \label{regimegeneraltSymNpR}
(1-2t)m +4t^2h-2t(1-t) l_{\rm NS} \leq 2t-1 \,,
\ee
This expression captures all states in the single pole regime for $t=2$. For $t>2$, we expect there to be $2^{t-1}-2$ additional regimes, coming from all the poles that lie between $p_L$ and $p_R$. One would obtain similar expression for the generating function and could write a piecewise single residue generating function with $2^{t-1}$ regimes.

An important point to mention is that all the states for which we have given exact expressions have both a weight and charge that scale with $m$ in the large $m$ limit. This means they are not light with respect to the lightness condition 1 given in \rref{lightness1}.
Since their weight scales with $m$ (which is proportional to $c$), they have Planckian energies and it is therefore a slight abuse to name them perturbative states. However, they are still polar states, which means they are below the black-hole threshold in AdS$_3$. One can still think about them as perturbative in some generalized sense: from \rref{tstandardSF} and \rref{tstandardSFpR} we infer that their growth is of the form \rref{eq:cS1} with $\alpha=1/2$, which still falls into the very sparse criteria.

\subsection{Fractional spectral flow}\label{ss:fracflow}

We have seen that in the previous section that identifying the vacuum with the most polar term that could be allowed by the index presents some puzzles. In such a theory, the vacuum and all states close to it would not contribute to the elliptic genus and the first non-zero terms are already at Planckian energies. We will now present a different physical interpretation to $\chi_m$. 

The most polar terms that appear in the symmetric product for our Siegel paramodular form are those of the form
\be
p^{ m} q^0 y^{ -m/t} \,,
\ee
and thus with discriminant $\Delta_{\rm min}=-m^2/t^2$.
We  want to map these states to the vacuum. To do this we spectral flow by a fractional amount. That is, we consider a spectral flow transformation that changes the charge $l$ in the following way
\begin{equation}
l\rightarrow l-m/t~,
\end{equation}
that is, with $\eta=-1/(2t)$. This ensures that the state with $l=m/t$ is mapped to a state with $l=0$ as expected for the vacuum. Since the polarity is bounded from below we have the inequality $\Delta\geq \Delta_{\text{min}}$. It follows from this that also $h$, the power of $q$, is bounded by
\begin{equation}
h\geq \frac{l^2}{4m}+\frac{\Delta_{\text{min}}}{4m}~.
\end{equation}
After the fractional spectral flow, the state with $l=0$ gives the lowest bound on $h$ as expected for the vacuum. This allows us to identify the central charge as the lowest value of $24 h$, that is,
\be
c= \frac{6 m}{t^2}~,
\ee
where $m$ is the index of the weak Jacobi form.  This contrasts the usual $c=6m$ for the Calabi-Yau sigma models we discussed in the previous subsection. Because the central charge has changed from the Calabi-Yau case, the charge vectors we are interested in will be different. After shifting by this vacuum energy, the spectral flow transformation is
\be
h=n+\frac{l}{2t}+\frac{m}{2t^2} \,, \qquad l_{\rm sf}=l+ \frac{m}{t} \,.
\ee
In terms of these new variables, the condition (\ref{regimepL}) to be in the $p_L$-residue regime becomes
\be\label{lightfractional}
-m +4t^2h  \leq -2\ .
\ee 
This is particularly appealing since it is an exact analogue of the lightness condition of the $t=1$ case in \eqref{singleresidue}.
Expressing the multiplicities (\ref{cLonepole}) in terms of our new NS variables \be
c_{\rm sf}(m,h,l_{\rm sf}):= c(m,n,l)=c\left(m,h-\frac{l_{\rm sf}}{2t},l_{\rm sf}-\frac{m}{t}\right)~,
\ee
gives
\bea
c_{\rm sf}^s(m,h,l_{\rm NS})&=&(\frac{m}{t}-2th) d_t(h-\frac{l_{\rm sf}}{2t})\tilde{d}_t(t(t h + \frac{l_{\rm sf}}{2})) \notag \\
&=&(\frac{m}{t}-2th) d_t(h-\frac{l_{\rm sf}}{2t})d_t(th+ \frac{l_{\rm sf}}{2})~.
\eea
This expression is valid and non-zero provided
\begin{align}
\frac{m}{t}-2th&\geq  1 ~, \cr
h-\frac{l_{\rm sf}}{2t} & \geq  - \frac{1}{t}~, \cr
t \, h+ \frac{l_{\rm sf}}{2} & \geq  - \frac{1}{t} \,.
\end{align}

The generating function of the $c_{\rm NS}^s$ is
\bea
Z_{\rm sf}^s(\tau,\rho,z) &=&\sum_{h} \sum_{l_{\rm sf}} \sum_{m\geq 2ht^2+t}  (\frac{m}{t}-2th) d_t(h-\frac{l_{\rm sf}}{2t})d_t(th+ \frac{l_{\rm sf}}{2}) q^h y^{l_{\rm sf}}p^m \,.
\eea
We can perform the sum over $m$ and find
\bea
Z_{\rm sf}^s(\tau,\rho,z) &=&\sum_{h\geq} \sum_{l_{\rm sf}}d_t(h-\frac{l_{\rm sf}}{2t})d_t(th+\frac{l_{\rm sf}}{2}) \left(qp^{2t^2}\right)^h y^{l_{\rm sf}} \sum_{m'\geq 0}   m' \left(p^t\right)^{m'} \notag \\
&=&\frac{p^t}{(1-p^t)^2} \sum_{h\geq} \sum_{l_{\rm sf}}d_t(h-\frac{l_{\rm sf}}{2t})d_t(th+\frac{l_{\rm sf}}{2}) \left(qp^{2t^2}\right)^h y^{l_{\rm sf}}  \notag \\
&=& \frac{p^t}{(1-p^t)^2} \sum_{r\geq-1/t} \sum_{s\geq -1/t} d(r)d(s) \left(qp^{2t^2}\right)^{\frac{tr+s}{2t}} y^{s-tr} \\
&=& \frac{p^t}{(1-p^t)^2}\frac{1}{\eta(\tau/2 + t^2 \rho - t z)^{24/t} \eta(\tau/2t+t\rho+z)^{24/t}}~.
\eea
Again, we further multiply the result by the spectrally flowed version of $p \phi_{k,1}$ to get the generating function of the symmetric product. We have
\be
p\, \phi_{k,1}= p q^{1/t}y(1-y^{-1})^2 \prod_{n\geq1} (1-q^n)^{24/t - 4}(1-q^n y)^2(1-q^n y^{-1})^2 \,.
\ee
We then do the spectral flow transformation
\be
y \to y q^{1/2t} \,, \qquad p\to p q^{1/2t^2} y^{1/t}
\ee
and obtain
\be
p\, \phi^{\rm sf}_{k,1}= pq^{1/2t+1/2t^2} y^{1/t-1}(1-yq^{1/2t})^2  \prod_{n\geq1} (1-q^n)^{24/t - 4}(1-q^{n+1/2t}y)^2(1-q^{n-1/2t}y^{-1})^2\ .
\ee
In total we find
\be\label{eq:qwer}
{\cal Z}_{\rm sf}^s=\frac{1}{(1-p^t)^2} \prod_{n\geq1} \frac{(1-q^n)^{24/t - 4}(1-q^{n-1+1/2t}y)^2(1-q^{n-1/2t}y^{-1})^2}{(1-q^{n/2}p^{t^2n}y^{-t n})^{24/t}(1-q^{n/2t}p^{t n}y^{ n})^{24/t}} \,.
\ee
This expression will give the exact Fourier coefficient provided we are in the regime
\be
 h  \leq \frac{m+2t-1}{4t^2}\ .
\ee
This time we find that the single pole regime is compatible with the lightness condition \rref{lightness1}. We therefore have many states with sub-Planckian energies and the formula above gives the generating function of the degeneracy of such states. 
Note that the $m\to\infty$ limit of this product can be obtained by extracting the coefficient of the $p=1$ pole  and setting $p=1$. We find
\be \label{ZinfSymNFracSF}
{\cal Z}^\infty_{\rm sf}= \prod_{n\geq1} \frac{(1-q^n)^{24/t - 4}(1-q^{n-1+1/2t}y)^2(1-q^{n-1/2t}y^{-1})^2}{(1-q^{n/2}y^{-t n})^{24/t}(1-q^{n/2t}y^{ n})^{24/t}} \,.
\ee
For $y=1$ this is simply a product of eta functions. The growth of the coefficients is thus clearly of supergravity type, i.e.\ of the form \eqref{eq:cS1} with $\alpha=1/2$, rather than Hagedorn, i.e.\ of the form \eqref{eq:cH1}.

\section{Supergravity interpretation}\label{sec:sugra}

We finally turn to the supergravity interpretation of the spectrum of light states of the symmetric product orbifold CFTs for our five examples. Our main findings in Sec.~\ref{sec:sfl} were the generating functionals of negative discriminant states that lie in the single pole regime. As we observed above, the single pole regime captures the closest states to the vacuum that contribute to the index, and hence it is our starting point to have a discussion on the gravitational features. 

In the following we will start with a review of the exact agreement among the KK spectrum of type IIB supergravity on ${\rm AdS}_3 \times S^3 \times K3$ with the spectrum of light BPS operators in the elliptic genus of $K3$. This corresponds to our example with $t=1$, and it serves as a guiding principle to what we expect for our remaining four examples. For $t>1$ we discuss the features and challenges to find a suitable supergravity dual to our counting formulas based on our findings in Sec. \ref{ss:halfint} and \ref{ss:fracflow}; we cover the half-integer  spectral flow in Sec. \ref{sec:cycomp}, and the fractional spectral flow in Sec. \ref{sec:orbads}. 

\subsection{Supergravity spectrum of ${\rm AdS}_3 \times S^3 \times K3$}
Let us briefly restate the supergravity results of \cite{deBoer:1998us}. That is, we assume that the supergravity spectrum is given by the KK reduced spectrum of  type IIB supergravity on AdS$_3 \times S^3 \times K3$ The spectrum decomposes into representations of the AdS supergroup $SU(1,1|2)_L \times SU(1,1|2)_R$. In the KK spectrum only short representations of $SU(1,1|2)$ appear, which we will denote by $(j)_S$; the short representation of both left and right movers is denoted $(j,j')_S$. The character $\chi_j(q,y)= \Tr_{(j)_S} (-1)^F q^{L_0}y^{J^3_0}$ of the representation $(j)_S$ is given by \cite{deBoer:1998us}
\bea
\chi_0(q,y)&=& 1~,\cr
\chi_1(q,y)&=& \frac{q^{1/2}}{(1-q)(y-y^{-1})}(y^2-y^{-2}-2q^{1/2}(y-y^{-1}))~,\cr
\chi_j(q,y)&=&  \frac{q^{j/2}}{(1-q)(y-y^{-1})} (y^{j+1}-y^{-j-1}-2 q^{1/2}(y^j-y^{-j})+q(y^{j-1}-y^{-j+1}))\ .
\eea
Note that $\chi_j(1,1)=1$. Following the prescription of \cite{deBoer:1998us}, we associate an additional degree $d(j,j')$ to representations $(j,j')_S$ of $SU(1,1|2)_L \times SU(1,1|2)_R$, with corresponding fugacity $p$. In total we write $(j,j';d)_S$ for such a short multiplet. The spectrum is then given by \cite{deBoer:1998us}\footnote{Note that there is a typo in (2.8) and also (5.9) in \cite{deBoer:1998us}.}
\be\label{sugraspectrum}
\bigoplus_{\hat m\geq 0} \bigoplus_{i,j} h^{i,j} (\hat m+i,\hat m+j;\hat m+1)
\ee
where $h^{i,j}$ are the Hodge numbers of $K3$, that is $h^{0,0}=h^{2,0}=h^{0,2}=h^{2,2}=1$ and $h^{1,1}=20$.

We want to count states that correspond to $|{\rm anything}\rangle_L\otimes |{\rm chiral ~primary}\rangle_R $, and for this we set  $\bar q = \bar y =1$ as we count the short representations  $(j,j';d)_S$.  To capture these states, it is convenient to first introduce a single-particle partition function $s(p,q,y)$ for the supergravity spectrum, which reads
\begin{align}\label{ssugra}
s(p,q,y)&=\sum_{m,n,l} c_{\rm sugra}(m,n,l) p^m q^n y^l \cr &= \sum_{m\geq 0} \sum_{i,j} h^{i,j}\chi_{m+i}(q,y)p^{m+1}
\\&=
\frac{1}{(1-q)(y-y^{-1})}\sum_{i,j}h^{i,j}p^{i+1}q^{i/2} \left(\frac{(y^{i+1}-2q^{1/2}y^{i}+qy^{i-1})}{1-pq^{1/2}y}
-\frac{(y^{-i-1}-2q^{1/2}y^{-i}+qy^{-i+1})}{1-pq^{1/2}y^{-1}}\notag
\right)
\end{align}
From this we can in principle extract the degeneracies of the single particle configurations, $c_{\rm sugra}(m,n,l)$, but we will refrain from doing so for the moment. Instead we want to look at the multi-particle spectrum, that is the second quantization of this. The generating function of this is the usual DMVV formula \cite{Dijkgraaf1997},
\be\label{2ndsugra}
{\cal Z}_{\rm sugra}=\prod_{m>0,n,l}\frac{1}{(1-p^m q^n y^l)^{c_{\rm sugra}(m,n,l)}}~.
\ee
An important point here is that the $c_{\rm sugra}(m,n,l)$ are essentially constant: they are bounded since the coefficients in $\chi_j$ are of order 1. This means that (\ref{2ndsugra}) is essentially a product of Dedekind-eta functions, which means that ${\cal Z}_{\rm sugra}$ has  growth of the form \eqref{eq:cS1} rather than Hagedorn growth.
 
Let us use our result in (\ref{ZSymNS}) for ${\cal Z}^s_{\rm NS}$ to recover the central result of 
\cite{deBoer:1998us}, that is that ${\cal Z}_{\rm NS}$ agrees with ${\cal Z}_{\rm sugra}$ provided $h \leq (m+1)/4$, which is exactly our condition in \eqref{lightNSK3}. 
The proof in  \cite{deBoer:1998us} involved to observe explicitly that even though the first quantized coefficents $c_{\rm sugra}$ and $c_{\rm NS}$ do not agree, their `first moments' do,
\begin{align}\label{sum}
\sum_m c_{\rm sugra}(m,n,l)=\sum_m c_{\rm NS}(m,n,l)~, \cr 
\sum_m mc_{\rm sugra}(m,n,l)=\sum_m mc_{\rm NS}(m,n,l)~ ,
\end{align}
which is enough to establish agreement of the second quantized partition function for light states. The advantage of (\ref{ZSymNS}) is that we can directly read off the single residue version of $c_{\rm NS}$,
\begin{align}
c^s_{\rm NS}(2,0,0)=2~, \quad c^s_{\rm cft}(0,n\geq 1,0)=-20~,\cr c^s_{\rm NS}(0,n-1/2,\pm1)=-2~,
\quad c^s_{\rm NS}(n,n/2,\pm n)=24~.
\end{align}
These then immediately agree with the sugra sums in (\ref{sum}).

\subsection{Compactifications of Calabi-Yau manifolds}\label{sec:cycomp}

We now turn to our expressions (\ref{tstandardSF}) and (\ref{tstandardSFpR}) for $t>1$, which are the generating functionals that represent the lowest states appearing in the NS sector.\footnote{As we mentioned round \eqref{regimegeneraltSymNpR} there are additional states in this sector that we have omitted for sake of simplicity. These omitted states behave in a similar fashion as  (\ref{tstandardSF}) and (\ref{tstandardSFpR}) for the purpose of the arguments in this section: their energies are Planckian. } At first sight they appear quite promising, as they give supergravity type growth just as for the $K3$ case. However, an important difference is that  none of the states are perturbative:  they do not obey the inequality $h\leq (m+1)/4$ (which we used for $t=1$) and they do not satisfy the lightness condition \rref{lightness1}. Still, the fact that their growth is not Hagedorn suggests that there may be a supergravity interpretation.

There is of course an obvious generalization of (\ref{sugraspectrum}): we can try to formally replace $K3$ by some higher dimensional Calabi-Yau manifold $M$, and use its Hodge numbers $h^{i,j}$. On the CFT side this is not an issue, since in that case we get a well-defined symmetric orbifold of a higher dimensional Calabi-Yau sigma-model.  On the gravitational side it is far from clear that this replacement will make sense physically. Formally we get the KK reduced spectrum on AdS$_3 \times S^3 \times M$, even though we have no right to expect a consistent supergravity theory on such a background. Still let us pursue this interpretation for the short time being.

The idea is to take  \eqref{sugraspectrum} and \eqref{ssugra} with the hodge numbers of $M$: this will lead to mathematically well-defined expressions for $c_{\rm sugra}$.  One could therefore hope to find Hodge numbers $h^{i,j}$ which give $c_{\rm sugra}$ that match $c_{\rm NS}$ extracted from, \eg, the $t=2$ SMF. 
Note that $s(p,q,y) = p\sum_{j}h^{0,j} + O(p)+O(q)$. This means that the term $p^1q^0y^0$ of ${\cal Z}_{\rm sugra}$ has coefficient $\sum_{j}h^{0,j}$. On the other hand it is straightforward to check that this state satisfies (\ref{regimegeneraltSymN}), but that there is no such term in (\ref{tstandardSF}), which implies that $\sum_{j}h^{0,j}=0$. There are however many non-vanishing terms in (\ref{tstandardSF}) such as $(m=4,h=1,l_{NS}=2)$ which in ${\cal Z}_{\rm sugra}$ are proportional to $\sum_{j}h^{0,j}$, which obviously contradicts our attempted matching. Maybe not surprisingly, this indicates that this interpretation is too naive.

In this argument, we associated the same degree $d(j,j')$ as in \cite{deBoer:1998us} to include $p$  in ${\cal Z}_{\rm sugra}$ for any Calabi-Yau manifold. It may be possible to counter our negative result in this subsection by introducing a different grading in $p$ to the supergravity spectrum. We don't have evidence that this will lead to a positive outcome, but we have not explored it in detail.  

Another alternative is to consider the KK spectrum of backgrounds of the form AdS$_3 \times S^2 \times M$. This approach will also not lead to a successful path for cases where we have a string/M-theory realisation. For example, the supergravity elliptic genera was studied in \cite{Gaiotto:2006ns,Kraus:2006nb}, and their results leads to  a growth of the perturbative spectrum of the form \eqref{eq:cS1} with $\alpha=2/3$. For better or worse, our examples have a significantly slower growth regardless of the spectral flow sector.  

\subsection{Fractional spectral flow and orbifolds of AdS$_3$}\label{sec:orbads}

Let us now turn to the interpretation of the generating functionals built by a fractional spectral flow: the generating functional \eqref{eq:qwer}. In this case, the counting formula captures perturbative states, and we have computed their degeneracy in the infinite central charge limit in \rref{ZinfSymNFracSF}. We would like to give a supergravity interpretation to these states. There are several odd features of the formula \rref{ZinfSymNFracSF} that makes challenging a bulk interpretation (and as a matter of fact, a CFT interpretation as well). First, the weights are no longer half-integer but rather fractionally quantized. Second, the formula is not invariant under $l_{\rm sf}\to-l_{\rm sf}$. We were not able to find a satisfying candidate for a gravity dual based solely on the formula \rref{ZinfSymNFracSF}, which we leave for future work. However, there are some directions that could unveil the putative gravity dual.

First, note that fractional spectral flows have been studied before in the context of orbifolds of AdS$_3$ \cite{Maldacena:2000dr,Cooper:1997qr,Martinec:2001cf,Son:2001qm}. From the worldsheet point of view the orbifold introduces twisted sectors that can be identified with the fractional spectral flow sectors \cite{Son:2001qm}.  In this context, is possible that the dual we are looking for is a singular $\mathbb{Z}_t$ orbifold of AdS, in which case  fractional quantization  and asymmetry between $l_{\rm sf}$ and $-l_{\rm sf}$ would be expected.\footnote{It is easy to show that a fractional spectral flow transformation acting on a parity invariant Jacobi form leads to an asymmetric spectrum. } At the moment we do not have a candidate gravitational theory. As we discussed in Sec. \ref{ss:fracflow}, the vacuum of the theory is in this fractional sector and we are not aware of theories with an orbifold of AdS$_3$ where this is the case. We hope that as new developments occur related to string compactifications of AdS$_3$, we will have more insight if this is a viable route. See for example \cite{CouzensMartelliSchafer-Nameki2017,CouzensLawrieMartelliEtAl2017} and references within.

Second, one could try to change variables such that the weights are no longer fractional. In some sense, this means we were using the wrong variable and one should simply replace $\tau$ by $t\tau$. Although this takes care of the fractional modding, it cannot be accomplished without changing the modular properties of the elliptic genus which we  would need to justify. A similar type of scenario occured in \cite{Datta:2017ert}, where the authors discuss the orbifolds AdS$_3\times (S^3\times T^4)/G$. In that case, it was the charges that were not appropriately quantized and so the chemical potential $z$ needed to be unwrapped, that is, rescaled.  One could imagine a situation where something of the sort needs to happen for $\tau$. This would mean that the elliptic genus would be related to a weak Jacobi form of a subgroup of $SL(2,\mathbb{Z})$ by an unwrapping procedure. The work of \cite{Datta:2017ert}  suggests that rescaling $\tau$ and orbifolding AdS$_3$ may be canonically related. It would be interesting to investigate this further.

%%%%%%%%%%%%%%%%%%%%%%%%%%%%%%%%%%%%%%%%%%%%%%%%%%%%%%%%%%%%%%%%%%
%%%%%%%%%%%%%%%%%%%%%%%%%%%%%%%%%%%%%%%%%%%%%%%%%%%%%%%%%%%%%%%%%%

\section{Discussion}\label{sec:discussion}

\subsection{Results}

In this paper, we presented a constructive algorithm to compute negative discriminant Fourier coefficients of the reciprocal of Siegel paramodular forms, $1/\Phi_k$, where $\Phi_k$ is obtained from an exponential lift of a weak Jacobi form. We focused on cases where $1/\Phi_k$ had second order poles dictated only by the Humbert surface $H_1(1)$. This gave five cases: the well-known Igusa cusp form $\Phi_{10}$, along with four other examples. We could then obtain the Fourier coefficients of negative discriminant states by a simple residue prescription around the poles of  $1/\Phi_k$; these residues are controlled in an elegant fashion by Dedekind-eta functions.

This methodology was then used to capture Fourier coefficients of symmetric product orbifold CFTs, with an emphasis on the limit of large central charge. The expressions were particularly simple for values of the charges where only a single pole contributed. We were particularly interested in sparseness or very sparseness of the Fourier coefficients, which indicates either a stringy dual or a more conventional  supergravity dual. In our examples the growth was always compatible with supergravity. This is a consequence of the form of the residues that capture the degeneracies: a finite number of Dedekind-eta functions have a sub-Hagedorn growth for large values of $h$, which leads to \eqref{eq:cS1} with $\alpha=1/2$.

To give a proper supergravity interpretation to the Fourier coefficients, we had to perform a spectral flow transformation. This step is important since the AdS vacuum as well as the perturbative supergravity states are usually the lightest states in the NS sector, whereas the Fourier coefficients of $1/\Phi_k$ and ${\cal Z}$ come from a Ramond sector elliptic genus. We suggested two choices for the spectral flow transformation, corresponding to two different interpretations of the central charge of the CFT. We discussed both possibilities, finding that each case had peculiarities.

In the standard half-integer spectral flow, we found that the contribution of all perturbative states cancelled and we were only left with states who have Planckian energy. The growth of such states is still compatible with supergravity although it is a slight abuse to name them perturbative states since their energy is Planckian. We also discussed a possible fractional spectral flow, finding that in that case there is a well-defined low energy perturbative spectrum. However, the states found this way were not charge conjugation invariant and had fractional weights.

Finally, we discussed possible supergravity interpretations for either scenarios. For the four new examples, we tried to compare the generating functions we found to a putative supergravity on AdS$_3 \times S^3 \times M $, for $M$ a Calabi-Yau manifold, and could not find a proper matching. We also discussed how the fractional spectral flows could correspond to orbifolds of AdS$_3$, but left a more precise investigation of this idea for future work. We now discuss some future directions that would be interesting to explore.

\subsection{Outlook}

\subsubsection*{Supergravity and CFT interpretation}

The biggest challenge in giving a physical interpretation to the counting formulas we derived is that we know neither the CFT nor the gravity theory. We have a family of weak Jacobi forms that we wish to interpret as the elliptic genera of a family of CFTs $\mathcal{C}_N$. We then want to find a supergravity theory that is dual to $\mathcal{C}_N$ and weakly coupled in the large $N$ limit. With only a counting formula in hand, it is quite challenging to proceed since different theories can have the same elliptic genera.

It is perhaps easier to start with the CFT side, since one can ask which two dimensional CFTs admit the weak Jacobi forms as their elliptic genera. A natural interpretation we discussed is to consider the elliptic genus of a higher dimensional Calabi-Yau Sigma model $M$. The family of CFTs in that case is simply
\be \label{symNdisc}
\mathcal{C}_N \equiv \frac{M^{\otimes N}}{S_N} \,.
\ee
We already saw that the issue with this interpretation is that the contribution of the vacuum and all light states vanish. We find a supergravity type growth, but it is only applicable for states with Planckian energy which is usually beyond the strict supergravity regime.

There is another issue with this interpretation. As discussed in the introduction, we  hope to discover a family of CFTs that are given by the symmetric product orbifolds of a seed theory $\mathcal{C}$, but only at weak coupling. We are hoping that these theories admit an exactly marginal deformation that lifts all non-protected states and leaves us with a supergravity theory at strong coupling. The elliptic genus captures only the supergravity states since it is protected and hence invariant under the marginal deformation. The issue is that the exactly marginal operator must couple the $N$ copies and is therefore necessarily in a twisted sector of \rref{symNdisc}. The lightest state of all non-trivial twisted sector is the ground state of the twist-2 sector with weights
\be
(h,\bar{h})=\left(\frac{ c}{16},\frac{ \bar{c}}{16}\right) \,.
\ee
For a Calabi-Yau d-fold, we have $c=\bar{c}=3d$. For $d>5$, this gives weights that are greater than one and there is therefore no hope of finding any exactly marginal operator. With this interpretation, there would still be hope to find exactly marginal deformations for some of our examples but not for all of them (for example not for $t=4$). Furthermore, a gravity dual of the type AdS$_3\times CY_d$ simply doesn't make sense in the framework of supergravity for $d>4$.

For the two reasons explained above, it seems appealing to look for a different CFT interpretation. This  is what led us to consider fractional spectral flows. The problem is that we don't yet have candidate CFTs with the appropriate central charges. It would be very interesting to build a candidate CFT, and we hope to return to this question in future work.

One could also try to give a direct supergravity interpretation to the counting formulas. Perhaps the peculiar form of the counting formulas (in particular the fractional weights and the unbalance between opposite charges) can help in identifying the relevant gravitational theory. We were unfortunately unable to do so for the moment. If it could be achieved, it would be very interesting to look for black hole solutions of those theories and investigate whether the black hole entropy is correctly accounted for by the method described in \cite{Belin:2016knb}. This would provide a highly non-trivial check for the new duality.

Finally, one interesting feature we noticed is that states satisfying the single pole regime introduced in Sec. \ref{sec:singlepole} are the relevant negative discriminant states to discuss the lightness condition \eqref{lightness1} for our examples. These negative discriminant states were also deduced  in \cite{Gomes:2015xcf} by exploring the black hole residue formula and the Rademacher expansion for CHL models. It would be interesting to complement these two methods.

\subsubsection*{Towards a complete classification of symmetric products}

In this paper, we only considered generating functionals whose poles were described by the Humbert surface $H_1(1)$. It would be very interesting to investigate the growth of Fourier coefficients for other exponential lifts. A generic weak Jacobi form will have other Humbert surfaces as well and one could hope to give a complete classification of the growth of symmetric products using our methodology. In principle, our method should be applicable to compute Fourier coefficients of other instances of $1/\Phi_k$ obtained by an exponential lift. There would be multiple tessellations of the Siegel upper half plane, each corresponding to a Humbert surface. One would need to carefully track the contribution of the residues as one crosses each pole but this can in principle be done.

The technical difficulty will be the form of the residues for other Humbert surfaces. The surface $H_1(1)$ always enables the mapping to the pole $z=0$ where the residue is simple, namely a product of two Dedekind-eta functions. For other surfaces, it will be more complicated. It is important to note that the form of the residues will dictate whether the growth is supergravity like or Hagedorn like. It would be very interesting to investigate this further. We would like to emphasize that there is hope for a complete classification here. At the end of the day, the SMFs are specified by very few elements: the polar terms of the seed theory. Each of these will give rise to a particular Humbert surface and all that is left to do is understand the residue at those poles. With this in hand, one can hope to find the full set of symmetric products whose duals are compatible with a supergravity.

Finally, note that one can also add zeros to the SMFs rather than poles. An example of this is the weak Jacobi form
\bea
\tilde{\phi}_{0,2}(\tau,z)& \equiv& -\frac{1}{24} \phi_{0,1}^2(\tau,z)-\frac{23}{24} \phi_{-2,1}^2(\tau,z)E_4(\tau) \notag \\
&=& -y^{-2}+3y^{-1} -10 +3y-y^2 + \ldots
\eea
This weak Jacobi form will lift to a SMF with poles at $H_1(1)$ of multiplicity two, but it will also have zeros along the Humbert surface $H_4(2)$. It would be interesting to understand the effect of the zeros as well.

To conclude, there are a finite number of possibilities (for fixed index) that one needs to explore and one can then formulate a complete classification of the growth of Fourier coefficients in symmetric products. It would be very interesting to perform this task and understand whether the examples considered in this paper are special from that point of view. Furthermore, it is important to note that the residues are also important in the context of logarithmic corrections to black hole entropy \cite{Belin:2016knb}. They will also determine how far the Cardy regime can be extended and are therefore of particular relevance. We hope to investigate this point further in the future.

\section*{Acknowledgements}
We are happy to thank Miranda Cheng, Shouvik Datta, Jan de Boer, Victor Godet, Valery Gritsenko, Sameer Murthy, Natalie Paquette, Erik Verlinde, Roberto Volpato and Max Zimet for helpful discussions. This work was initiated at the Aspen Center for Physics, which is supported by National Science Foundation grant PHY-1607761. AB would like to thank the String Theory group at ETH for hospitality. AB is supported by the NWO VENI grant 680-47-464 / 4114. AC and JG are supported by Nederlandse Organisatie voor Wetenschappelijk Onderzoek (NWO) via a Vidi grant. CAK is supported by the Swiss National Science Foundation through the NCCR SwissMAP. This work is supported by the Delta ITP consortium, a program of the Netherlands Organisation for Scientific Research (NWO) that is funded by the Dutch Ministry of Education, Culture and Science (OCW).

%%%%%%%%%%%%%%%%%%%%%%%%%%%%%%%%%%%%%%%%%%%%%%%%%%%%%%%%%%%%%%%%%%
%%%%%%%%%%%%%%%%%%%%%%%%%%%%%%%%%%%%%%%%%%%%%%%%%%%%%%%%%%%%%%%%%%

\appendix

\section{Review of the black hole contour}\label{app:bh}

In this appendix we briefly discuss the contour used for positive discriminant states, i.e. $\Delta >0$, which we loosely associate with black holes in the gravitational side. This is a succinct overview of the detailed discussion in \cite{Sen:2007qy}, generalized to our five examples here.  This review is useful to contrast the choices and techniques in this case relative to those states with negative discriminant studied in the main text.  

First, we recall that the Siegel upper half plane is given by
\be
\text{Im} \tau >0 \,, \qquad \text{Im} \rho >0 \,, \qquad \text{Im} \tau\text{Im} \rho - (\text{Im} z)^2>0 \,.
\ee
As for $\Delta<0$ states, for positive discriminant  we also need to pick what expansion we are doing in $y$. We will expand around $y=0$ and  select $l<0$. The convergence of the expansion then implies 
\be
\text{Im} z >0  \,.
\ee
The basic characteristic of the contour is to set
\be
\text{Im} \tau \gg 1 \,, \qquad \text{Im} \rho \gg 1 \,, \qquad \text{Im} z \gg 1 \,, \qquad \text{Im} \tau\text{Im} \rho - (\text{Im} z)^2 \gg 1 \,,
\ee
with the range for the real parts being
\be
0 \leq \text{Re} \tau, \text{Re} \rho, \text{Re} z <1 \,.
\ee
Note that this is not a closed contour in $\H_2$. The strategy explained in \cite{Sen:2007qy}  is to close the contour by adding a segment along the same real parts but with
\be
\text{Im} \tau \sim 1 \,, \qquad \text{Im} \rho \sim 1 \,, \qquad \text{Im} z \sim 1
\ee
This segment will give a small contribution compared to the residue picked up at the poles inside where the contribution is exponentially big. For example, when the only divisor is the Humbert surface $H_1(1)$, the dominant pole was given by
\be
t(\tau \rho - z^2)+z=0 \,,
\ee
which gives a contribution \cite{Belin:2016knb}
\be\label{eq:ccardy}
c(\Delta)\approx e^{\pi \sqrt{\Delta}/t}~,
\ee
which is much bigger than the contribution from the surface that closes the contour.

\section{A comment on symmetric products of very special weak Jacobi-Forms}
\label{app:veryspecial}
In this section, we review the difference between the weak Jacobi forms that lead to the generating functions $1/\Phi_k$ whose only pole is $H_1(1)$ and those called \textit{very special weak Jacobi forms} in \cite{Benjamin:2015vkc}.

The NS partition function of a very special weak Jacobi forms at large $c$  is chacterized by the fact that its "perturbative" states, by which here we mean states below black hole formation, have Planckian energy. This is what defines them to be special. In other words, in the NS sector we have
\begin{equation}
\phi_{NS}(q,y)=q^{-\frac{m}{4}+a}+\ldots~,
\end{equation}
where $a$ is $\mathcal{O}(m/4)$. Symmetric products of such functions where shown in \cite{Benjamin:2015vkc} to give rise to a sub-Hagedorn growth of Fourier coefficients upon specialization to $z=0$, and therefore amenable to a possible supergravity interpretation. Eventhough our  four new examples have similar properties, we show in this section that the conclusions of \cite{Benjamin:2015vkc} do not apply here since our four examples are not of the very special type.

The weak Jacobi form under study, which has index $t$, has a most negative discriminant
\begin{equation}\label{eq:amin1}
\Delta_{\rm min}=-1~,
\end{equation}
in contrast with the usual weak Jacobi forms which allow for $\Delta_{\rm min}=-t^2$. As usual, we can write the Jacobi form in a theta function expansion as 
\begin{equation}
\phi_{w,t}(\tau,z)=\sum h_{\mu}(\tau)\theta_{\mu,t}(\tau,z)~,
\end{equation}
where $w$ is the weight which is zero in this case, and
\begin{equation}
\theta_{\mu,t}(\tau,z)=\sum_{l=\mu\,\text{mod}(2t)}q^{\frac{l^2}{4t}}y^l~,
\end{equation}
is the Jacobi-theta function. On the other hand,
\begin{equation}
h_{\mu}(\tau)=q^{-\frac{\mu^2}{4t}}\sum_{n=0}^{\infty}d_{\mu}(n)q^{n}~,
\end{equation}
is a vector-valued modular form. More details about the theta-function decomposition can be found in \cite{9781468491647}.

For each sector $\mu$, the terms in $h_{\mu}(\tau)$ with negative powers of $q$ define the polar terms, that is, all the terms for which $-\mu^2/4t+n$ is negative. This means that in each sector $\mu$, the polarities must obey  the condition
\begin{equation}\label{cond max polar}
0>-\frac{\mu^2}{4t}+n\geq -\frac{1}{4t},\;n\geq 0~,
\end{equation}
because of \eqref{eq:amin1}. Consider the sector with $\mu=t$. By (\ref{cond max polar}) we have the following constraint on $n$
\begin{equation}
\frac{t}{4}>n\geq \frac{t}{4}-\frac{1}{4t}~.
\end{equation}
We want to show that there is no solution for $n$ provided that $t>1$. To do this, write $t=p+4q$ with both $p,q$ positive integers and $0\leq p<4$. Plugging this in the inequality above, we find  
\begin{equation}
q+\frac{p}{4}>n\geq q+\frac{p}{4}-\frac{1}{4t}~.
\end{equation}
For $p>0$ we always have $1>p/4-1/4t>0$ and so there is no integer in the interval $[q+p/4-1/4t,q+p/4[$. And similarly for $p=0$ because the interval is open on the right side.
Therefore, we conclude that there {\it is no polar term in the sector $\mu=t$}.

After a half spectral flow transformation the theta functions mix between themselves, but the polarities are preserved. In the NS sector we still have the decomposition
\begin{equation}\label{NS very special}
\phi_{0,t}^{NS}(\tau,z)=\sum_{\mu}\tilde{h}_{\mu}(\tau)\theta_{\mu+t,t}(\tau,z)
\end{equation}
but now the sector $\mu=t$ is mapped to the sector $\mu=2t\sim 0$. Since there was no polar term in this sector, we see that the NS sector does not contain a vacuum. 

We can also show that all terms in (\ref{NS very special}) have non-negative powers of $q$. Writing $\tilde{h}_{\mu}(\tau)$ as
\begin{equation}
\tilde{h}_{\mu}(\tau)=q^{-\frac{\mu^2}{4t}}\sum_{n=0}^{\infty}g_{\mu}(n)q^{n}
\end{equation}
we need to show that
\begin{equation}
n-\frac{\mu^2}{4t}+\frac{(\mu-t)^2}{4t}\geq 0,\;\;n\geq 0
\end{equation}
where $(\mu-t)^2/4t$ comes from the theta function, and $\mu\geq 0$ and $n-\mu^2/4t<0$. The case with $\mu=t$ was already analysed. Since the polarity always obeys the lower bound $-1/4t$ we must have
\begin{equation}
n-\frac{\mu^2}{4t}+\frac{(\mu-t)^2}{4t}\geq -\frac{1}{4t}+\frac{(\mu-t)^2}{4t}.
\end{equation}
But $(\mu-t)^2\geq 1$, and thus the RHS is always non-negative as we wanted to show.

This exercise therefore implies that if we set $y=1$ in $\phi^{NS}_{0,t}(q,y)$ we obtain a modular form with Fourier expansion of only positive powers of $q$, that is, it must be a cusp form of zero weight. But this is not possible because there is no modular invariant cusp form and so $\phi^{NS}_{0,t}(q,y=1)$ must vanish identically. The very special Jacobi forms studied in \cite{Benjamin:2015vkc} are clearly not of this type.

\section{The elliptic genus of a Calabi-Yau \label{calabiyau}}
Take a non-linear sigma-model of some Calabi-Yau $d$-fold $M$. Its elliptic genus is then given by some weak Jacobi form $\varphi(\tau,z)$. 
What does the condition of its exponential lift only having divisors at $H_1(1)$ mean?
For this let us discuss simply the contribution of 1/2-BPS primary states to the elliptic genus, that is states that are in short representations for both the left- and rightmovers. Their multiplicities are given by the Hodge numbers $h^{i,j}$. For a CY $d$-fold, the $N=2$ superconformal algebra has central charge $3d$. In the Ramond sector it has $d+1$ short representations of $U(1)$ charge $Q=0,1,\ldots,d$, whose characters are given by $\chi_Q(\tau,z)$. Their Witten index is given by $\chi_Q(\tau,0)=(-1)^Q$. In the NS sector these flow to the chiral primaries. The 1/2-BPS contribution to the RR partition function is given by
\be
Z_{1/2}(\tau,\bar\tau) = \sum_{i,j=0}^d h^{i,j} \chi_i(\tau,z)\chi_j(\bar\tau,\bar z)\ .
\ee
To get the contribution to the elliptic genus, we specialize $\bar z =0$ to get
\be\label{eq:zzz2}
Z_{1/2}(\tau,z) = \sum_{i=0}^d \chi_i(\tau,z) \sum_{j=0}^d h^{i,j}(-1)^j\ .
\ee
We see that depending on the Hodge numbers, there can indeed be cancellations, such that not all left-moving short representations contribute to the elliptic genus. Often for a `proper' Calabi-Yau we assume that
\begin{align}
&h^{0,0}=h^{0,d}=1\ , \cr
&h^{0,i}= 0~, \qquad {\rm for}~~ i=1,\ldots, d-1\ .
\end{align}
This is equivalent to assuming that the theory has no enhanced symmetry, i.e. that its symmetry algebra is given by the $N=2$ superconformal algebra together with the holomorphic $(0,d)$-form, which together form the Odake algebra. In such cases we see that $\chi_0(\tau,z)$ contributes with multiplicity $1+(-1)^d$, which in particular means that the vacuum does not contribute for odd Calabi-Yaus, as we saw already for CY 3-folds.

 If $h^{0,j}>0$, then that means that there is an additional spin $j/2$ symmetry in the theory. This is the case if the holonomy group is not $SU(d)$, but a strict subgroup thereof. From \eqref{eq:zzz2} it is clear that if $h^{0,i}$ does not vanish there may be additional cancellations: a familiar example is $T^4$, where the contribution of the vacuum and the $(0,2)$ form is cancelled by the fermions. (In fact, the entire elliptic genus vanishes in that case.)

This shows that it is indeed possible that the vacuum does not appear in the elliptic genus. This happens generically for odd Calabi-Yaus, but for even Calabi-Yaus it can only happen if the theory has an enhanced symmetry.

\bibliographystyle{JHEP-2}
\bibliography{ref}

\end{document}